\documentclass[12pt,notoc]{JHEP3}

\usepackage{amsmath,amssymb,euscript,array,cite,mathrsfs}

\def\oldf{f}

\newcommand{\startappendix}{
\setcounter{section}{0}
\renewcommand{\thesection}{\Alph{section}}}
\newcommand{\Appendix}[1]{
\refstepcounter{section}
\begin{flushleft}
{\large\bf Appendix \thesection: #1}
\end{flushleft}}
\usepackage{epsfig}

\newcommand{\Tr}{\operatorname{Tr}}

\def\Bv{{\boldsymbol v}}

\def\Bvarpi{{\boldsymbol \varpi}}

\def\B0{{\boldsymbol 0}}
\def\BF{{\boldsymbol F}}
\def\BK{{\boldsymbol K}}
\def\BX{{\boldsymbol X}}

\def\BOmega{{\boldsymbol\Omega}}
\def\Bvarpi{{\boldsymbol\varpi}}
\def\Bpi{{\boldsymbol\pi}}
\def\CP{{\mathbb C}P}

\def\Tr{{\rm Tr}}

\def\det{{\rm det}}

\def\Z{{\bf Z}}

\def\Z{\boldsymbol{Z}}

\newcommand{\BH}{\boldsymbol{H}}

\newcommand{\Be}{\boldsymbol{e}}

\def\Dbarslash{\,\,{\raise.15ex\hbox{/}\mkern-12mu {\bar D}}}
\def\Dslash{\,\,{\raise.15ex\hbox{/}\mkern-12mu D}}
\def\delslash{\,\,{\raise.15ex\hbox{/}\mkern-9mu \partial}}
\def\delbarslash{\,\,{\raise.15ex\hbox{/}\mkern-9mu {\bar\partial}}}

\def\LAG{\mathscr{L}}

\def\II{\mathscr{I}}

\newcommand{\MAT}[1]{\begin{pmatrix} #1\end{pmatrix}}

\newcommand{\EQ}[1]{\begin{equation}\begin{split} #1
\end{split}\end{equation}}
\newcommand{\AL}[1]{\begin{subequations}\begin{align} #1
\end{align}\end{subequations}}
\newcommand{\SP}[1]{\begin{equation}\begin{split} #1
\end{split}\end{equation}}

\title{Classical and Quantum Solitons in the Symmetric Space Sine-Gordon Theories}
\author{Timothy J. Hollowood\\
Department of Physics,\\ University of Wales Swansea,\\
Swansea, SA2 8PP, UK.\\
E-mail: \email{t.hollowood@swansea.ac.uk}}
\author{and J.~Luis Miramontes\\
Departamento de F\'\i sica de Part\'\i culas and IGFAE,\\
Universidad
de Santiago de Compostela\\ 15782 Santiago de Compostela, Spain\\
E-mail: \email{jluis.miramontes@usc.es}}

\abstract{We construct the soliton solutions in the symmetric space
  sine-Gordon theories. The latter are a series of integrable field
  theories in $1+1$-dimensions which are associated to a symmetric
  space $F/G$, and are related via the Pohlmeyer reduction to
  theories of strings moving on symmetric spaces. We show that the solitons are
  kinks that carry an internal moduli space that can be
  identified with a particular co-adjoint orbit of the unbroken subgroup
  $H\subset G$. Classically the solitons come in a continuous spectrum which encompasses the perturbative fluctuations of the theory as the kink charge becomes small. We show that the solitons can be
  quantized by allowing the collective coordinates to be
  time-dependent to yield a form of quantum mechanics on the
  co-adjoint orbit. The quantum states correspond
to symmetric tensor representations of the symmetry group $H$ and  have the interpretation of a fuzzy geometric version of the co-adjoint
  orbit. The quantized finite tower of soliton states includes the perturbative modes at the base.}

\begin{document}

\section{Introduction}

The Symmetric Space sine-Gordon (SSSG) theories are a large class of relativistic integrable field theories in $1+1$ dimensions that generalize the sine-Gordon and complex sine-Gordon theories. They 
arise as the result of imposing the Pohlmeyer reduction on a sigma model with a symmetric space $F/G$ as the target space~\cite{Pohlmeyer:1975nb} (for a recent review see~\cite{Miramontes:2008wt} and references therein), and can be described as the gauged WZW model for a coset $G/H$ deformed by a particular potential term.
Remarkably, the SSSG theories have been shown to be equivalent, at least classically, to the world-sheet theory for strings propagating on the symmetric space~\cite{Tseytlin:2003ii,Mikhailov:2005qv,Mikhailov:2005sy}, and since the basic building blocks for the geometry of the AdS/CFT correspondence are symmetric spaces like $AdS_n=SO(2,n-1)/SO(1,n-1)$, $S^n=SO(n+1)/SO(n)$ and
$\CP^n=SU(n+1)/U(n)$ it is clearly of interest to understand them at the quantum level.
Moreover, it has been suggested that the SSSG theory which is classically equivalent to superstrings on $AdS_5\times S^5$ is also equivalent at the quantum level~\cite{Grigoriev:2007bu,Mikhailov:2007xr} (see also~\cite{Grigoriev:2008jq,Roiban:2009vh,Hoare:2009rq,Hoare:2009fs,Iwashita:2010tg,Hoare:2010fb}).
That case involves a generalization of the SSSG theories involving supergroups
where symmetric spaces are generalized to semi-symmetric spaces~\cite{Zarembo:2010sg}.

In the present work, we shall consider the SSSG theories for ordinary compact groups, involving Type I symmetric spaces, and develop general techniques for quantizing them. The approach adopted is a generalization of the well-known technique for quantizing the sine-Gordon theory itself \cite{Zamolodchikov:1978xm}. The idea is to focus on the soliton solutions that are known to exist in these theories. Although these solitons have been constructed elsewhere \cite{Hollowood:2009tw,Hollowood:2009sc}, here we develop a variant of the dressing method that is intrinsic to the SSSG theories in which relativistic covariance is manifest. We show that the solitons are kinks, as in the sine-Gordon theory. The new ingredient is that, classically, there is a continuous moduli space of solutions on which the kink charge varies smoothly. The internal moduli space has the form of a (co-)adjoint orbit of $H\subset G$. On the other hand, we argue that the perturbative excitations themselves carry charges under the global subgroup of the gauge group $H$, and that this charge is itself a kink charge. It follows from this remarkable feature that there is a limit where the solitons become the perturbative fluctuations of the theory.  

We then describe how the solitons can be semi-classically quantized by allowing the internal collective coordinates to vary with time leading to an effective quantum mechanical description. This effective description leads to a quantization of the co-adjoint orbit and gives rise to a tower of states transforming in symmetric tensor representations of $H$.  For the $\CP^n=SU(n+1)/U(n)$ example, the resulting semi-classical spectrum matches precisely the spectrum of the S-matrix conjectured  in~\cite{Hollowood:2010rv}.

The approach that we adopt it somewhat complementary to the one adopted in a series of papers \cite{Hoare:2009fs,Hoare:2010fb} based on a perturbative expansion of the theory. Ultimately, in our picture, the perturbative fields are simply the solitons with small kink charge that lie at the base of a tower of states, and the two approaches for constructing the S-matrix should agree for these states. This problem will be addressed elsewhere.

The whole formalism that we develop can be generalized to the supergroup (or semi-symmetric space) case that is needed in the full AdS/CFT correspondence, and it is described in a companion paper \cite{us2}.

The plan of the paper is as follows. In Section~\ref{PohlmeyerR}, we review the construction of the SSSG theories associated to a symmetric space $F/G$, whose Lagrangian formulation is provided by the gauged WZW action for a coset $G/H$ deformed by a potential. In particular, we show that the Noether charge corresponding to global gauge transformations is a kink charge. In Section~\ref{pmass}, we describe the spectrum of perturbative fluctuations, showing that the perturbative modes are kink-like solutions with non-trivial boundary conditions. We also discuss the quantization of these modes at tree level, which leads to the quantization of their kink/Noether charge. In Section~\ref{Inth}, we use the Lax form of the equations of motion of the SSSG theories to uncover the infinite tower of conserved charges that follow from integrability. We write them in terms of a subtracted monodromy matrix whose form is specialized to the case of the solutions obtained via the dressing transformation method. Then, in Section~\ref{Dressing}, we construct the soliton solutions by means of the dressing method. They are kinks with an internal moduli space that can be identified with a particular (co-)adjoint orbit of the symmetry group $H\subset G$. The close relationship between the spectrum of solitons and the spectrum of perturbative modes is made explicit and manifests the fact that the latter are solitons with very small kink charge. In Section~\ref{SMquantum} we quantize the solitons in the semi-classical approximation using a well-known technique that originated with Manton \cite{Manton:1981mp}. This leads to towers of states transforming in the symmetric tensor representations of the symmetry group $H$, which can be seen as a fuzzy geometric version of the (co-)adjoint orbit~\cite{Balachandran:2005ew}. Finally, Section~\ref{Discussion} contains our conclusions and a discussion about the sigma model interpretation of the SSSG solitons. There is one Appendix.
 
\section{The Symmetric Space Sine-Gordon Theories}
\label{PohlmeyerR}

In this section, we review the construction of the SSSG theories.
A more detailed discussion can be found in~\cite{Miramontes:2008wt} and references therein. The starting point is a symmetric space which can be realized as a
quotient of two Lie groups $F/G$. The group in the numerator $F$ admits 
an involution $\sigma_-$ whose stabilizer is the
subgroup $G$. Acting on the Lie algebra of $F$, the involution gives rise to the canonical 
decomposition
\EQ{
{\mathfrak f} = {\mathfrak g} \oplus {\mathfrak p}
\quad \text{with} \quad 
[{\mathfrak g},{\mathfrak g}]\subset {\mathfrak g}\>, 
\quad [{\mathfrak g},{\mathfrak p}]\subset 
{\mathfrak p}\>, 
\quad [{\mathfrak p},{\mathfrak p}]\subset {\mathfrak g}\>,
\label{CanonicalDec}
}
where ${\mathfrak g}$ and ${\mathfrak p}$ are the $+1$ and $-1$ eigenspaces of $
\sigma_-$, respectively.
In this paper we will only consider compact symmetric spaces of Type I, which are those for which $F$ is a compact simple Lie group. These are listed in Table~\ref{GTable}, which includes Cartan's notation that we shall use for brevity.

\TABLE[ht]{
{\small\begin{tabular}{cccc}
\hline\hline
\\[-10pt]
Cartan & $F/G$ &  Involutions & $H_\text{regular}$ 
\\
\\[-10pt]
\hline
\\[-10pt]
AIII & ${SU(n+p)}/{S(U(n)\times U(p))}$ & $\sigma_-(U)=I_{np}UI_{np}$ &$S(U(n-p)\times U(1)^{p})$\\[7pt]
BDI & ${SO(n+p)}/{SO(n)\times SO(p)}$ &$\sigma_+(U)=U^*$ &$SO(n-p)$ \\
&&$\sigma_-(U)=I_{np}UI_{np}$ &\\[7pt]
CII & ${Sp(n+p)}/{Sp(n)\times Sp(p)}$ & $\sigma_+(U)=J_{n+p}U^*J_{n+p}^{-1}$& $Sp(n-p)\times Sp(1)^p$\\ &&$\sigma_-(U)=K_{np}UK_{np}$&\\[7pt]
AI &${SU(n)}/{SO(n)}$ &$\sigma_-(U)=U^*$& $\emptyset$\\[7pt]
AII & ${SU(2n)}/{Sp(n)}$ & $\sigma_-(U)=J_n U^* J_n^{-1}$&$SU(2)^n$\\[7pt]
DIII&${SO(2n)}/{U(n)}$ &$\sigma_+(U)=U^*$&  $SU(2)^{\frac n2}$\\
&&$\sigma_-(U)=J_nUJ_n^{-1}$&$SU(2)^{\frac{n-1}2}\times U(1)$\\[7pt]
CI&${Sp(n)}/{U(n)}$ &$\sigma_+(U)=J_nU^*J_n^{-1}$& $\emptyset$\\ && $\sigma_-(U)=J_nUJ_n^{-1}$ &\\[7pt]
\hline\hline
\\[-5pt]
\end{tabular}}
\label{GTable}
\caption{\small The Type I symmetric spaces corresponding to the classical groups including Cartan's classification, the associated involutions, and the subgroup $H=H_\text{regular}$ for the regular choice of $\Lambda$. The two expressions for $H$ for the case DIII correspond to $n$ even and odd, respectively. We choose $n\geq p$.}
}

The SSGG equations are formulated at the level of the Lie algebra $\mathfrak f$
and involve a group field $\gamma(t,x)\in G\subset F$. They take the zero-curvature form\footnote{In our notation, $x^\pm  = t\pm x$ are light-cone coordinates, and for a general 2-vector we use $a_\pm=\frac12(a_0\pm a_1)$. Our choice of metric is $\eta=\text{diag}(1,-1)$ and we normalize the anti-symmetric symbol with $\epsilon_{01}=1$. In order to compare with the notation of \cite{Hoare:2009fs}, they have $x^\pm  =\frac{1}{\sqrt{2}}( t\pm x)$ and $a_\pm=\frac{1}{\sqrt{2}}(a_0\pm a_1)$. In addition, our element $\Lambda=\Lambda_\pm$ is equal to $\frac{1}{\sqrt{2}}\mu T$, and their $k$ equals our $\kappa/(2\pi)$ in~\eqref{gWZW}.}
\EQ{
\big[\partial_++\gamma^{-1}\partial_+\gamma+\gamma^{-1}A_+^{(L)}\gamma-\Lambda_+\>,
\;
\partial_-+A_-^{(R)}-\gamma^{-1}\Lambda_-\gamma\big]=0\>.
\label{www1}
}
Here, $\Lambda_+$ and $\Lambda_-$ are constant elements of a maximal abelian subspace of  ${\mathfrak p}$, the~$-1$ eigenspace of the Lie algebra of $F$. The dimension of the maximal abelian subspaces of ${\mathfrak p}$ defines the rank of the symmetric space. Therefore, if $\text{rank}(F/G)=1$, which includes the case of the spheres $S^n$ and the complex projective spaces $\CP^n$, we can always fix $\Lambda_+=\Lambda_-\equiv \Lambda$ without loss of generality. In this paper we will consider generic symmetric spaces with $\text{rank}(F/G)\geq 1$ but, for simplicity, we will keep the restriction $\Lambda_+=\Lambda_-\equiv \Lambda$. The more general case with $\Lambda_+\not=\Lambda_-$ will be discussed elsewhere (but see the comment at the end of section~\ref{pmass}).

A central role will be played by the subgroup $H\subset G$ that keeps $\Lambda$ fixed under adjoint action; namely, $\Lambda = U\Lambda U^{-1}$ 
for $U\in H$. The Lie algebra of $H$, ${\mathfrak h}$, consists of the elements of ${\mathfrak g}$ that commute with $\Lambda$. It is important in what follows that the adjoint action of 
 $\Lambda$ gives rise to an orthogonal decomposition
\EQ{
{\mathfrak f} ={\mathfrak f}^\perp\oplus{\mathfrak f}^\parallel\ ,\qquad
{\mathfrak f}^\perp=
 \mathop{\rm Ker} {\mathop{\rm Ad}}_{\Lambda}\ , \qquad{\mathfrak f}^\parallel = \mathop{\rm Im} {\mathop{\rm Ad}}_{\Lambda}\ ,
\label{Orthogonal}
}
which is always true provided that $F$ is compact.\footnote{In contrast, for $F$ non-compact it is not generally true that any $\Lambda$ gives rise to~\eqref{Orthogonal}. An explicit example involving $F=SO(2,n)$ can be found in\cite{Miramontes:2008wt} (Sec. 5.3).
}
Schematically, this decomposition satisfies
\SP{
[{\mathfrak f}^\perp,{\mathfrak f}^\perp]\subset {\mathfrak f}^\perp\ ,\qquad
[{\mathfrak f}^\perp,{\mathfrak f}^\parallel]\subset{\mathfrak f}^\parallel,
\label{Orthogonal2}
}
and it is worth noticing that ${\mathfrak h}={\mathfrak g}^\perp={\mathfrak f}^\perp\cap{\mathfrak g}$.

The fact that, for $F$ compact, any $\Lambda\in\mathfrak{f}$ give rise to an orthogonal decomposition of the form~\eqref{Orthogonal} can be proved in two steps. First, let us consider the usual realization of $\mathfrak{f}$ in terms of a Chevalley basis of its complexification, which consists of Cartan generators $H^a$, with $a=1,\ldots, \text{rank}(\mathfrak{f})$, and step operators $E_{\pm\alpha}$, where $\alpha$ is a positive root. In particular, they satisfy
$[H^a,H^b]=0$ and $[H^a,E_{\pm\alpha}]=\pm\alpha^a E_{\pm\alpha}$. This provides the following anti-Hermitian basis for the compact Lie algebra $\mathfrak{f}$: $
t^a= i H^a$,  $t^\alpha= E_\alpha -E_{-\alpha}$, and $t^{\overline\alpha}= i(E_\alpha +E_{-\alpha})$, and it can be explicitly checked that any linear combination $\Lambda=\sum_a \mu_a t^a$ of the Cartan generators gives rise to~\eqref{Orthogonal}.
Next, we will show that for any $\Lambda\in \mathfrak{f}$ there exists a group element $\overline\varphi\in F$ such that $\overline\varphi^{-1}\Lambda \overline\varphi$ is a linear combination of Cartan generators. In order to do that, 
let us consider the (maximal) abelian subalgebra $\mathfrak{s}\subset \mathfrak{f}$ spanned by the generators $t^a$. It can be shown that $\mathfrak{s}$ contains a (regular) element $k_0$ whose centraliser in $\mathfrak{f}$ is $\mathfrak{s}$; namely, 
$\mathfrak{s}=\{k\in \mathfrak{f}\;/\; [k_0,k]=0\}$. Then, $\varphi\to \Tr\big(\Lambda \varphi k_0\varphi^{-1})$ defines a continuous function on the compact group $F$ and, therefore, it takes a minimum for, say, $\varphi=\overline{\varphi}$. For each $T\in \mathfrak{f}$, this requires that
\SP{
0=\frac{d}{ds} \Tr\big(\Lambda \overline\varphi e^{sT} k_0 e^{-sT}\overline\varphi^{-1})\big|_{s=0}=\Tr\big(T\,[k_0,\overline\varphi^{-1}\Lambda \overline\varphi]\big)\,.
}
Since the trace form is non-degenerate, this implies that $[k_0,\overline\varphi^{-1}\Lambda \overline\varphi]=0$ and, thus, $\overline\varphi^{-1}\Lambda \overline\varphi\in\mathfrak{s}$, which ensures that~\eqref{Orthogonal} is satisfied for any $\Lambda\in\mathfrak{f}$.

Clearly the subgroup $H$ depends on $\Lambda$.
In the generic case, $\Lambda$ is a ``regular'' element of ${\mathfrak p}$, which means that ${\mathfrak p}^\perp$ is a maximal abelian subspace. Therefore, its dimension is the rank of the symmetric space, and the dimension of $H= H_\text{regular}$ is minimal.  Then, for symmetric spaces with $\text{rank}(F/G)>1$ there are non-generic choices of $\Lambda$ for which $H\supset H_\text{regular}$ changes discontinuously. For the most part we shall assume the regular case, and we will see that the non-generic choices correspond to different SSSG theories since they involve gauging a different group. Table \ref{LTable} lists the generic expressions for $\Lambda$, up to conjugation, in the defining representation of $F$. Those expressions involve a basis $E_{ab}$ with $(E_{ab})_{ij}=\delta_{ai}\delta_{bj}$ for $N\times N$ matrices. Later we will need the associated vectors $\Be_a$ 
with $E_{ab}\Be_c=\delta_{bc}\Be_a$.\footnote{Vectors of the defining representation will be denoted in boldface.}

\TABLE[ht]{
{\small\begin{tabular}{ccc}
\hline\hline
\\[-10pt]
Cartan &  Rank &$\Lambda$
\\
\\[-10pt]
\hline
\\[-10pt]
AIII & $p$& $\sum_{a=1}^p m_a\big(E_{n+a,n-p+a}-E_{n-p+a,n+a}\big)$   \\[7pt]
DBI & $p$& $\sum_{a=1}^p m_a\big(E_{n+a,n-p+a}-E_{n-p+a,n+a}\big)$   \\[7pt]
CII & $p$& $\sum_{a=1}^p m_a\big(E_{n+a,n-p+a}-E_{n-p+a,n+a}-E_{2n+p+a,2n+a}
+E_{2n+a,2n+p+a}\big)$ \\ [7pt]
AI & $n-1$& $i\sum_{a=1}^nm_aE_{aa}$ \\[7pt]
AII &$n-1$&$i\sum_{a=1}^nm_a\big(E_{aa}+E_{a+n,a+n}\big)$  \\[7pt]
DIII &  $[n/2]$ &$\sum_{a=1}^{[n/2]}m_a\big(E_{a+1,a}-E_{a,a+1}-E_{a+n+1,a+n}+E_{a+n,a+n+1}\big)$\\[7pt]
CI & $n$ &$i\sum_{a=1}^nm_a\big(E_{aa}-E_{a+n,a+n}\big)$\\[7pt]
\hline\hline
\\[-5pt]
\end{tabular}}
\label{LTable}
\caption{\small The algebra element $\Lambda$, up to conjugation, in the defining representation for the Type~I symmetric spaces associated to the classical groups. In the cases AI and AII, we also have the constraint $\sum_{a=1}^nm_a=0$. The rank of the symmetric space is also included. We choose $n\geq p$.}
}

The quantities $A^{(L)}_+$ and 
$A^{(R)}_-$ in~\eqref{www1} can be interpreted as light-cone components of gauge fields
associated to a $H_L\times H_R$ gauge symmetry under which
\EQ{
\gamma\longrightarrow U_L\gamma U_R^{-1}\ ,\qquad U_{L/R}\in H\ ,
\label{gaugeLR1}
}
and 
\EQ{
A_-^{(R)}\longrightarrow U_R\big(A_-^{(R)}+\partial_-\big)U_R^{-1}\ ,\qquad
A_+^{(L)}\longrightarrow U_L\big(A_+^{(L)}+\partial_+\big)U_L^{-1}\ .
\label{gaugeLR2}
}
A Lagrangian formalism can be found by identifying $
  A_-\equiv A_-^{(R)}$ and $A_+\equiv A_+^{(L)}$ as the two light-cone
components of a gauge field, and by imposing the constraints~\cite{Bakas:1995bm}
\SP{
\Big(\gamma^{\mp1}\partial_\pm\gamma^{\pm1}
+\gamma^{\mp1}A_\pm\gamma^{\pm1}\Big)^\perp=A_\pm\ .
\label{gco2}
}
These conditions
can be viewed as a set of partial gauge fixing
conditions~\cite{Grigoriev:2007bu,Miramontes:2008wt}.
They reduce the $H_L\times H_R$ gauge symmetry~\eqref{gaugeLR1} to the
$H$ vector subgroup\footnote{
Note that it is also possible to gauge
the axial or the vector subgroup of any overall $U(1)$ factor of
$H$ whilst still gauging the vector subgroup of the non-abelian
component. This gives rise to different Lagrangian
formulations of the theory~\cite{Miramontes:2008wt,Hollowood:2009tw} which are  related by a kind of T-duality \cite{Miramontes:2004dr}.
}
\EQ{
\gamma\longrightarrow U\gamma U^{-1}\>, \qquad U\in H\>,
\label{gaugeH}
}
under which $A_\mu$ transforms as a gauge connection:
\EQ{
A_\mu\longrightarrow U\big(A_\mu+\partial_\mu\big)U^{-1}\ .
\label{gaugeHA}
}
The partially gauge fixed equations-of-motion are then
\EQ{
\big[\partial_++\gamma^{-1}\partial_+\gamma
+\gamma^{-1}A_+\gamma,\partial_-+A_-\big]
=-[\Lambda,\gamma^{-1}\Lambda\gamma]
\label{eom3}
}
and these follow as the 
equations-of-motion of the action
\SP{
S&=S_\text{gWZW}[\gamma,A_\mu]-\frac \kappa{\pi}\int d^2x\,\Tr\left(\Lambda
\gamma^{-1}\Lambda\gamma\right)\ .
\label{ala}
}
Here, $S_\text{gWZW}[\gamma,A_\mu]$ is the usual gauged WZW action for $G/H$,
\EQ{
S_\text{gWZW}[\gamma,A_\mu]&=-\frac\kappa{2\pi}\int d^2x\,\Tr\,\Big[
\gamma^{-1}\partial_+\gamma\,\gamma^{-1}\partial_-\gamma+2A_+\partial_-\gamma\gamma^{-1}\\ &~~~~~~~~~
-2A_-\gamma^{-1}\partial_+\gamma-2\gamma^{-1}A_+\gamma A_-+2A_+A_-\Big]
\\ &~~~~~~~~~+\frac{\kappa}{12\pi}\int d^3x\,\epsilon^{abc}\Tr\,\Big[\gamma^{-1}\partial_a\gamma\,
\gamma^{-1}\partial_b\gamma\,\gamma^{-1}\partial_c\gamma\Big]\ .
\label{gWZW}
}
So one interpretation of the theory is as the
gauged WZW model for $G/H$ deformed by 
the particular operator $\Tr\,(\Lambda\gamma^{-1}\Lambda\gamma)$.\footnote{This interpretation naturally leads to the description of the SSSG theories as integrable perturbations of coset CFTs. Some aspects of this description, including their quantum integrability and the calculation of the conformal dimension of the perturbing operator, have been discussed in~\cite{CastroAlvaredo:2000kq}.}
Notice that the partial gauge-fixing constraints \eqref{gco2} naturally arise as
the equations-of-motion of the gauge connection. 
The coupling of the
theory is the quantity $\kappa$ which is equal to
\EQ{
\kappa=\begin{cases} k & \text{unitary}\\ k/2 & \text{orthogonal, symplectic}\ ,\end{cases}
\label{Level}
}
where $k$ is the 
level of the WZW action which in the quantum theory must be a positive integer. The level plays the r\^ole of $\hbar^{-1}$ and, in particular, the classical limit corresponds to $k\gg1$.
Clearly, the theories with higher rank and with inequivalent choices for $\Lambda$ and, thus different groups $H$, are not deformable into one another.

One of the subtleties in the SSSG
theories is that the fields are non-trivial at $x=\pm\infty$ and so the WZ term and its variation requires careful treatment. In particular, it cannot strictly speaking be defined as an integral over a three-dimensional space with the two-dimensional spacetime as a boundary. One way to unambiguously define the action is, as in~\cite{Hoare:2009fs}, to use the condition of gauge invariance to pin down the expansion of the WZ term in terms of $\phi$. As explained in Appendix~A, this prescription involves the addition of 
the following $A_\mu$--dependent topological term to the na\"\i ve expression:
\EQ{
-\frac{\kappa}{2\pi} \int d^2x\> \epsilon^{\mu\nu}\partial_\mu \Tr\big(A_\nu \phi \big)\ ,\quad\text{with}\quad \gamma = e^\phi\ .
\label{Topol}
}

The issue of how to deal with the gauge symmetry is also central to understanding these theories.
As usual in a gauge theory, gauge fixing has the effect of removing unphysical degrees-of-freedom associated to local gauge transformations. However, global gauge transformations do not represent unphysical degrees-of-freedom and remain as symmetries of the gauge fixed theory~\cite{Coleman}. 

Ultimately, the $H$ gauge symmetry must be fixed in order to define a consistent quantum theory. At the level of the equations-of-motion one can fix the gauge by setting $A_\mu=0$, which follows from the fact that $A_\mu$ is a flat gauge connection on-shell,
\EQ{
[\partial_\mu+A_\mu,\partial_\nu+A_\nu]\approx 0\ .
\label{flat}
}
Off-shell, however, it is convenient to use a light-cone gauge which imposes the Lorentz invariant condition \cite{Hoare:2009fs,Hoare:2010fb}\footnote{In earlier work \cite{Hollowood:2009tw}, we have fixed the symmetry by using it to rotate the group field $\gamma$ to a chosen gauge slice. However, this kind of ``unitary gauge" is not consistent at the quantum level because the gauge orbits degenerate at certain points on the gauge slice, {\it e.g.\/}~$\gamma=1$. 
}
\EQ{
A_+=0\ .
}
At the level of the functional integral, the other component $A_-$ can be integrated out to yield the constraint
\EQ{
\big(\gamma^{-1}\partial_+\gamma\big)^\perp=0\ .
\label{gcond2}
}
In most of the rest of the paper, we will be almost exclusively interested in solutions of the equations-of-motion, and in that case we will always take solutions for which $A_\mu=0$, which means that from \eqref{gcond} we have the on-shell constraints
\EQ{
\big(\gamma^{-1}\partial_+\gamma\big)^\perp=\big(\gamma\partial_-\gamma^{-1}\big)^\perp=0\ .
\label{gcond3}
}
It is worth noticing that our solutions are also solutions to the equations of motion of the gauged fixed action of~\cite{Hoare:2009fs,Hoare:2010fb}.

For $A_\mu=0$, the potential term in \eqref{ala} has a space of minima given by constant group elements $\gamma\in H\subset G$. Classically, we would expect the choice of vacua at $x=\pm\infty$ to break the gauge symmetry and lead to a Higgs effect. However, in $1+1$-dimensions there is no symmetry breaking,
and one has to integrate over the flat directions of the potential of the field $\gamma$ at infinity in the functional integral.  In other words, when looking for classical solutions of the theory, we should allow for arbitrary values of $\gamma(\pm\infty)\in H$. Integrating over the possible values of $\gamma(\pm\infty)$ has the effect of restoring the $H$ symmetry, and the quantum states come in representations of the global part of the gauge group $H$. One way to understand why one is forced to integrate over the flat directions of the potential at infinity is that there exist kink excitations of arbitrarily small energy that can change the value of the field at infinity. These kinks are crucial to our story and we will find that classically they have a continuous spectrum without a mass gap. Quantum mechanically, the spectrum becomes quantized and a mass gap is generated.
In addition, it is worth pointing out that in the theories that we are discussing the gauge field is not dynamical and so does not confine in the IR, which confirms the expectation that physical states will carry global $H$ gauge charge, and that is exactly what we find in the following. From this perspective, it is important to notice that the residual gauge transformations left by the on-shell gauge fixing conditions $A_\mu=0$ are the global gauge transformations.\footnote{It is worth pointing out that the behaviour here is very different from a typical $G/H$ coset sigma model, where $H$ acts by right multiplication. In this case the $H$ gauge symmetry confines and the spectrum only consists of colour singlets (this is explained in Coleman's book \cite{Coleman}).}

The constraints \eqref{gco2} have the interpretation of the vanishing, on-shell, of what is na\"\i vely the Noether current corresponding to (global) gauge transformations\footnote{Where necessary, the notation $\approx$ will indicate equality on-shell.}
\SP{
J_\pm=\big(\gamma^{\mp1}\partial_\pm\gamma^{\pm1}+\gamma^{\mp1}A_\pm\gamma^{\pm1} -A_\pm\big)^\perp\approx0\ .
\label{gcond}
}
In fact this is just an example of the theorem of Hilbert and Noether that the current associated to a local symmetry vanishes on-shell, but crucially up to a total divergence which determines the charges of physical states. In particular, this means that the charge is determined by the behaviour of the fields at spatial infinity.\footnote{For a discussion of these issues in a modern context see, for example, \cite{Julia:1998ys,Silva:1998ii,Julia:2000er} and references therein.} In the present context, the total divergence is of the form $\epsilon^{\mu\nu}\partial_\nu \Phi$ which obviously cannot be deduced from the equations-of-motion since $\partial_\mu(\epsilon^{\mu\nu}\partial_\nu \Phi)=0$ identically. However, such a topological current is needed to account for the Noether charge of on-shell field configurations. The full expression for the Noether current including the topological contribution, which we denote ${\cal J}^\mu$, is obtained in Appendix A; writing the field as $\gamma=e^\phi$, we have (eq.~\eqref{ncor})
\EQ{
{\cal J}^\mu=J^\mu+\epsilon^{\mu\nu}\partial_\nu\phi^\perp\ .
\label{kss}
}
The topological term, which is generated by~\eqref{Topol}, is sensitive to the behaviour of the field at spatial infinity; indeed, on-shell $J^\mu=0$, and the Noether charge emerges as a kink charge:
\EQ{
{\cal Q}=\int dx\, {\cal J}^0\approx-\int dx\,\partial_1\phi^\perp=
-\phi^\perp(\infty)+\phi^\perp(-\infty)=q_0\in{\mathfrak h}\ .
\label{chgs1}
}
Note that at $x=\pm\infty$ the group field must lie in a minimum of the potential so that $\phi(\pm\infty)\in\mathfrak h$ and, consequently, the projection onto $\mathfrak h=\mathfrak g^\perp$ is unnecessary. 
Assuming that $\gamma(-\infty)$ and $\gamma(+\infty)$ commute, which will be true for the configurations that we consider, it follows that the kink charge is precisely
\EQ{
\gamma^{-1}(\infty)\gamma(-\infty)=e^{q_0}\ .
\label{chgs2}
}
The conclusion is that physical configurations that carry $H$ charge are actually kink-like configurations. In section \eqref{pmass}, we shall see how these charges emerge in the perturbative expansion and we shall verify the formula \eqref{chgs1} to lowest order.
An important observation is that the kink charge transforms under global gauge transformations as follows
\EQ{
q_0 \longrightarrow U q_0 U^{-1}\ , \qquad U\in H\ .
\label{AdjAction}
}
In other words, the kink charge takes values in (co-)adjoint orbits of $H$,\footnote{For compact semi-simple Lie groups the adjoint orbits are the same as the co-adjoint orbits.} which confirms that the states of these theories come in representations of the global part of the gauge group.

\noindent
{\bf The involutions}

It is useful to take a point of view where the group $F$ is thought of as a subgroup of $SU(N)$, where $N$ is the dimension of the defining representation of $F$~\cite{Harnad:1983we}. In examples for which $F$ is not the unitary group itself, the subgroup $F\subset SU(N)$ is then picked out as the invariant subgroup of an involution $\sigma_+$. 
In addition, we also have the involution $\sigma_-$  which leads to the 
decomposition $\mathfrak f=\mathfrak g\oplus\mathfrak p$ described in section \ref{PohlmeyerR}. These two involutions commute and, taken together, they generate a discrete group $\II$, which is either an ${\mathbb Z}_2$, for the unitarity cases where $\sigma_+$ is not needed, or an ${\mathbb Z}_2\times{\mathbb Z}_2$ group. We will denote the order of $\II$ as $n_0=2$ or 4, respectively. In the defining representation of $F$, the involutions are either holomorphic or anti-holomorphic; that is, for $U\in F$, of type
\SP{
\sigma_\text{hol}(U)=\theta U\theta^{-1}\qquad\text{or}\qquad
\sigma_\text{anti-hol}(U)=\theta 
U^*\theta^{-1}\ .
\label{lsi}
}
These involutions act on the (anti-hermitian) generators of the algebra as
\SP{
\sigma_\text{hol}(a)=\theta a\theta^{-1}\ ,\qquad\sigma_\text{anti-hol}(a)=\theta 
a^*\theta^{-1}\ .
\label{lsi2}
}
For the Type I symmetric spaces,  $\theta$ can be
either $I_{pq}$, $J_n$, or $K_{pq}$, where
\EQ{
I_{np}=\MAT{{\mathbb I}_n&0\\ 0&-{\mathbb I}_p}\ ,\qquad
J_n=\MAT{0&{\mathbb I}_n\\ -{\mathbb I}_n&0}\ ,\qquad
K_{np}=\MAT{{\mathbb I}_n&0&0&0 \\0 & -{\mathbb I}_p&0&0\\
0&0&{\mathbb I}_n&0\\ 0&0&0&-{\mathbb I}_p}\ .
}
It is useful to extend the action of $\II$ to vectors of the defining representation as
\EQ{
\sigma_\text{hol}(\Bvarpi)=\theta\Bvarpi\ ,\qquad\sigma_\text{anti-hol}(\Bvarpi)=\theta\Bvarpi^*\ .
}
The involutions for each case are listed in Table~\ref{GTable}.

\noindent
{\bf The Eigenvectors of $\Lambda$}

In the following, the 
eigenvectors and the eigenvalues of $\Lambda$ will play a central r\^ole. Since $\Lambda$ is anti-hermitian, the eigenvalues are imaginary:
\EQ{
\Lambda\Bv_a=im_a\Bv_a\ ,\qquad\Bv_a^\ast\cdot\Bv_b=\delta_{ab}\ .
}
Moreover, since $\sigma_\pm(\Lambda)=\pm\Lambda$,  the involutions act as permutations of the eigenvectors and eigenvalues:
\EQ{
\Lambda\sigma_\pm(\Bv_a)=\begin{cases}\pm im_a\sigma_\pm(\Bv_a) & \text{holomorphic}\\ \mp im_a\sigma_\pm(\Bv_a)& \text{anti-holomorphic}\ .\end{cases}
}
The above defines the action of the involutions on the eigenvalues, $\sigma_\pm(m_a)$, in an obvious way.

Notice that the eigenvalues $m_a$ parameterize the different possible choices of $\Lambda$. Therefore, the maximal number of linearly independent non-vanishing eigenvalues equals the rank of the symmetric space. 
In the SSSG theories, those eigenvalues are adjustable parameters, very similar to the adjustable mass scales of the homogeneous sine-Gordon theories~\cite{FernandezPousa:1996hi,FernandezPousa:1997zb,Miramontes:1999hx,CastroAlvaredo:1999em,Dorey:2004qc}.

Any element $\phi\in{\mathfrak g}$ can be expanded in terms of the eigenvectors as follows\footnote{Notice that $\tfrac14(1+\sigma_-) (1+\sigma_+)=\frac1{n_0} \sum_{\sigma\in\II}\sigma$ is a projector onto ${\mathfrak g}$.}
\EQ{
\phi=\sum_{\sigma\in\II}\sigma\Big(\phi_{ab}\>\Bv_a\Bv_b^\dagger\Big)\ ,\qquad \phi_{ab}=-\phi_{ba}^\ast\ .
\label{modes}
}
In particular, $\mathfrak g^\parallel$ is obtained by restricting the sum over $a,b$ to pairs of eigenvectors such that $m_a\not= m_b$, while  $\mathfrak h\equiv\mathfrak g^\perp$ is obtained by taking degenerate eigenvectors with $m_a=m_b$. Below we consider some of the cases in more detail. In all of them $\Lambda$ is assumed to be regular.

\noindent
{\bf(AIII)} The complex Grassmannians $SU(n+p)/S(U(n)\times U(p))$, with $n\geq p$. The subalgebras $SU(n)$ and $SU(p)$ are 
identified with the spaces spanned by $\Be_a$ for $1\leq a\leq n$ and $n+1\leq a\leq n+p$, respectively, or schematically
\EQ{
\left( \begin{array}{c|c}
SU(n) & 0 \\ \hline 
0&SU(p)
\end{array} \right)\ .
}
From Table~\ref{LTable}, the generic  form of  $\Lambda$ is
\EQ{
\Lambda=\sum_{a=1}^p m_a\big(E_{n+a,n-p+a}-E_{n-p+a,n+a}\big)=
\left( \begin{array}{cc|cc|cc}
\phantom{0000}&&  & && \\ 
 \phantom{0000}&&  & && \\ \hline 
 && & &-m_1& \\
 &&&&&\ddots\\ \hline
 && m_1&&&\\
 &&& \ddots& &
\end{array} \right)
\label{lam1}
}
where, in the final schematic form, the diagonal blocks are of size $(n-p)^2$, $p^2$ and $p^2$. The regular case corresponds to $m_a\not=0$ for any $a$, and $m_a\not= m_b$ for $a\not=b$. Then, the non-null eigenvalues are not degenerate, and their eigenvectors come in pairs
\EQ{
\Lambda\Bv_a^\pm=\pm im_a\Bv_a^\pm\ ,\qquad\Bv_a^\pm=\frac1{\sqrt{2}}\big(\Be_{n-p+a}\pm i\Be_{n+a}\big)\ , \qquad a=1,\ldots, p\ .
\label{eige}
}
For the $\sigma_-$ involution, we have
\EQ{
\sigma_-(\Bv_a^\pm)=I_{np}\Bv^\pm_a=\Bv_a^\mp\ .
 \label{swop}
}
In contrast, the null eigenvalue is $n-p$ times degenerate. A basis of eigenvectors is provided by $\Be_a$ for $a=1,\ldots,n-p$, and we have
\EQ{
\sigma_-(\Be_a)=I_{np}\Be_a=\Be_a\ .
}
Therefore, a generic null eigenvector is a linear combination of the form $\BOmega=\sum_{a=1}^{n-p} c_a \Be_a$ with complex coefficients, and
\EQ{
\sigma_-(\BOmega)=\BOmega\ .
}
In the following, $\BOmega$ will always denote a generic null eigenvector.

\noindent
{\bf(BDI)} The $\Lambda$ matrix for the real Grassmannians $SO(n+p)/SO(n)\times SO(p)$ is identical to the complex Grassmannians, and so the eigenvectors and the action of $\sigma_-$ is identical. For the $\sigma_+$ involution, which in this case is anti-holomorphic, we have
\EQ{
\sigma_+(\Bv_a^\pm)=\Bv_a^\mp\ ,\qquad\sigma_+(\BOmega)=\BOmega^\ast\ .
\label{InvRealG}
}

\noindent
{\bf(CII)} The ``quaternionic Grassmannians" $Sp(n+p)/Sp(n)\times Sp(p)$. In this case, from Table~\ref{LTable},
\EQ{
\Lambda=\left(\begin{array}{c|c}\Lambda'&0\\ \hline 0&-\Lambda'\end{array}\right)\ ,
}
where $\Lambda'$ is the same as \eqref{lam1}. 
Then, in the regular case, the non-null eigenvalues are two times degenerate, and their eigenvectors come in groups of four
\SP{
\Lambda \Bv_a^{\pm} = \pm i m_a \Bv_a^{\pm}, \qquad
\Lambda J_{n+p}\Bv_a^{\mp} = \pm i m_a J_{n+p}\Bv_a^{\mp}\ , \qquad 
a=1,\ldots, p\ ,
}
where $\Bv_a^\pm$ is as in \eqref{eige}.
The null eigenvalue is $2(n-p)$ times degenerate, and a basis of eigenvectors is provided by the pairs $\Be_a$ and $J_{n+p}\Be_a=-\Be_{n+p+a}$, $a=1,\ldots,n-p$.

\noindent
{\bf(AII)} $SU(2n)/Sp(n)$. In this case, from Table~\ref{LTable},
\EQ{
\Lambda=i\sum_{a=1}^n m_a\big(E_{aa}+E_{a+n,a+n}\big)
=i\left( \begin{array}{cc|cc}
m_1&&&\\ &\ddots&&\\ \hline &&m_1&\\ &&&\ddots
\end{array} \right)\ , \qquad \sum_{a=1}^n m_a=0\ ,
}
and the regular case corresponds to $m_a\not= m_b$ for $a\not=b$. Then, all the eigenvalues are two times degenerate, and their
eigenvectors come in pairs
\EQ{
\Lambda \Be_a = im_a \Be_a, \qquad \Lambda J_n\Be_a= im_a J_n\Be_a\ ,\qquad a=1,\ldots, n-1\ ,
}
where $J_n\Be_a=- \Be_{n+a}$. Therefore, the generic eigenvectors are linear combinations of the form
\EQ{
\BOmega_a = \alpha_a \Be_a + \beta_a \Be_{n+a}\ ,\qquad |\alpha_a|^2+|\beta_a|^2 =1\ ,
\label{EigenVAII}
}
where $\alpha_a$ and $\beta_a$ are complex coefficients. Then,
\EQ{
\sigma_-(\BOmega_a)= J_n \BOmega_a^\ast = \beta_a^\ast \Be_a - \alpha_a^\ast \Be_{n+a}\ .
}

\section{The Perturbative Excitations}
\label{pmass}

In this section, we describe the spectrum of perturbative fluctuations. Taking the on-shell gauge $A_\mu=0$, we expand the field $\gamma=e^\phi$ as
\EQ{
\gamma=1+\phi+\cdots\ ,
}
and then the  linearized equation-of-motion
is simply the free wave equation
\EQ{
\square\phi=\big(\partial_0^2-\partial_1^2)\phi =4\big[\Lambda,[\Lambda,\phi]\big]\ .
}
The constraints \eqref{gcond3} have the effect of removing the massless modes 
$\phi^\perp\in{\mathfrak h}$. In order to see this, decompose 
\EQ{
\phi=\phi^\perp+\phi^\parallel
}
and solve the constraints \eqref{gcond3} for $\phi^\perp$ order-by-order in the fluctuation $\phi^\parallel$. To lowest order,
\EQ{
\partial_\pm\phi^\perp=\pm\frac12[\phi^\parallel,\partial_\pm\phi^\parallel]^\perp+\cdots\ .
\label{wer}
}
Hence, to linear order $\phi^\perp$ is constant. 
However, it is interesting that to quadratic order
$\phi^\perp$ becomes non-vanishing. Pursuing this further we find that $\phi^\perp$ actually has a kink-like behaviour; to quadratic order,
\EQ{
\phi^\perp(x=\infty)-\phi^\perp(x=-\infty)=\int_{-\infty}^\infty dx\, \partial_1
\phi^\perp=
\frac12\int_{-\infty}^\infty dx\,
[\phi^\parallel,\partial_0\phi^\parallel]^\perp\ .
}
Remarkably, the right-hand side is minus the $H$ charge of a perturbative mode. In order to see this, note that the tree-level action for the perturbative modes is
\EQ{
S[\phi^\parallel]=-\frac \kappa{8\pi}\int d^2x\,\Tr\Big(\partial_\mu\phi^\parallel\partial^\mu\phi^\parallel-4[\Lambda,\phi^\parallel]^2+\cdots\Big)\ .
}
At the level of this action for the physical modes, the gauge symmetry becomes a global symmetry, 
\EQ{
\phi^\parallel\longrightarrow U\phi^\parallel U^{-1}\ ,
}
with the associated Noether current taking the form
\EQ{
{\cal J}_\mu=-\frac12[\phi^\parallel,\partial_\mu\phi^\parallel]^\perp\ .
}
Consequently, as anticipated in \eqref{chgs1} and \eqref{chgs2}, the Noether charge is equal to the kink charge
\EQ{
{\cal Q}=\int_{-\infty}^\infty dx\,{\cal J}^0=-\phi^\perp(x=\infty)+\phi^\perp(x=-\infty)\equiv q_0\ .
\label{NoetherKink}
}
It is worth emphasizing that something remarkable and surprising has happened: we think of the perturbative modes as describing localized wave-packets in space after suitable smearing in momentum space, and this is indeed true for $\phi^\parallel$; however, the full group field $\gamma$ is actually a kink-like solution with $\gamma(x=\infty)\neq\gamma(x=-\infty)$ due to the behaviour of $\phi^\perp$.

We can now go on to quantize the perturbative modes at tree level,
and this will lead to a quantization of the kink charge.
To be more specific, let us analyse the modes corresponding to the field
\EQ{
\phi^\parallel=\sum_{\sigma\in\II}\sigma\Big(\varphi\>\Bv_a\Bv_b^\dagger- \varphi^\ast\>\Bv_b\Bv_a^\dagger\Big)\ ,
\label{PertField}
}
where $\Bv_a$ and $\Bv_b$ are two fixed eigenvectors with eigenvalues $m_a\not= m_b$.
It satisfies
\EQ{
\big[\Lambda,[\Lambda,\phi^\parallel ]\big] =- (m_a-m_b)^2 \phi^\parallel\ ,
}
which shows that those modes are associated to particle states with mass
\EQ{
M=2|m_a-m_b|\ .
}
We can expand the complex field $\varphi$ in terms of on-shell modes in the form
\EQ{
\varphi=\int \frac{dp}{2\pi \sqrt{E(p)}}\>\Big(A(p)\>e^{-i(E(p)t-px)}+B^\dagger(p)\> e^{+i(E(p)t-px)}\Big)\ ,
\label{modes2}
}
where $E(p)=\sqrt{M^2+p^2}$ is the energy of the mode. 
Assuming that $m_b\not= -m_a$, quantization of $\phi^\parallel$ leads to the commutation relations
\EQ{
\big[\hat A(p),\hat A^\dagger(q)\big]=\big[\hat B(p), \hat B^\dagger(q)\big] =\frac{2\pi^2}{k}\> \delta(p-q)\ .
}
In the above, we have used the fact that
\EQ{
n_0\kappa=2k\ ,
\label{kapk}
}
which can be easily checked by looking at eq.~\eqref{Level} and Table~\ref{GTable}.
Then, a Fock space can be built up in the standard way, where $\hat A^\dagger(p)$ and $B^\dagger (p)$ create particles and anti-particles, respectively, of mass $M=2|m_a-m_b|$ with momentum $p$. 

It is a simple matter to calculate the $H$ charge of the state $\hat A^\dagger(p)|0\rangle$:
\EQ{
{\cal Q}=\frac{\pi}{k}\sum_{\sigma\in\II}\sigma(i\Bv_a\Bv_a^\dagger-i\Bv_b\Bv_b^\dagger)\ ,
\label{NoetherCharge}
}
which clearly commutes with $\Lambda$.
Hence, according to~\eqref{NoetherKink}, the kink charge is\footnote{It is important in this expression that the factor of $i$ cannot be moved out of the bracket, since for anti-holomorphic involutions $\sigma(ia)=-i\sigma(a)$.}
\EQ{
q_0= -\Delta\phi^\perp=\frac{\pi}{k}\sum_{\sigma\in\II}\sigma(i\Bv_a\Bv_a^\dagger-i\Bv_b\Bv_b^\dagger)\ .
}
Correspondingly, the charge of the state $\hat B^\dagger(p)|0\rangle$ is $-{\cal Q}$.
Since the perturbative field $\phi^\parallel$ transforms under global gauge transformations, the perturbative modes come in representations of $H$ which, in general, will be complex. Consider the field configuration $\phi^{\parallel\ast}$. The charge of the corresponding $A$-states is
\EQ{
\frac{\pi}{k}\sum_{\sigma\in\II}\sigma(i\Bv_b^\ast\Bv_b^t-i\Bv_a^\ast\Bv_a^t)= {\cal Q}^*\ .
\label{NoetherChargeB}
}
Therefore, if ${\cal Q}^\ast\not= {\cal Q}$ the representation is indeed complex, and the particles created by $\hat A^\dagger(p)$ and $\hat B^\dagger(p)$ transform in complex conjutate representations.
We refer to this situation as ``complex'. In contrast,  we will refer to the cases with ${\cal Q}^\ast={\cal Q}$ as ``real''

Finally, let us address the modes associated to eigenvalues with $m_b =-m_a$. Then, at least when $\Lambda$ is regular,  there exists an involution $\tilde\sigma\in\II$ such that
\EQ{
\Bv_b= \tilde\sigma(\Bv_a)\ .
}
Making the change $\sigma\to \sigma\tilde\sigma$, this leads to
\EQ{
\phi^\parallel=\sum_{\sigma\in\II}\sigma\Big(\varphi\>\Bv_a\Bv_b^\dagger- \varphi^\ast\>\Bv_b\Bv_a^\dagger\Big)=-\sum_{\sigma\in\II}\sigma\Big(\tilde\varphi^\ast\>\Bv_a\Bv_b^\dagger- \tilde\varphi\>\Bv_b\Bv_a^\dagger\Big)\ ,
\label{RealInv}
}
where $\tilde\varphi=\varphi$, or $\varphi^\ast$, if $\tilde\sigma$ is holomorphic, or anti-holomorphic, respectively.
Therefore, if $\tilde\sigma$ is anti-holomorphic then $\phi^\parallel=0$ and there is no perturbative mode associated to this choice of $\Bv_a$ and $\Bv_b$. Correspondingly, if  $\tilde\sigma$ is holomorphic then the field is invariant under the interchange of~$a$ and~$b$, its charge vanishes, and the mode is real.

\TABLE[ht]{
{\small\begin{tabular}{cccc}
\hline\hline
\\[-10pt]
Eigenvectors & Mass & Abelian Charges & $SU(n-p)$ Rep.\\
\\[-10pt]
\hline
\\[-10pt]
$(\Bv^\pm_a,\Bv^\pm_b)$ & $2|m_a-m_b|$ & $\pm(\Be_{n-p+a}+\Be_{n+a}-\Be_{n-p+b}-\Be_{n+b})$ & $[1]$\\[7pt]
$(\Bv^\pm_a,\Bv^\mp_b),\;\; a\not=b$ & $2|m_a+m_b|$ & $\pm(\Be_{n-p+a}+\Be_{n+a}-\Be_{n-p+b}-\Be_{n+b})$& $[1]$\\[7pt]
$(\Bv^\pm_a,\BOmega)$ & $2|m_a|$ & $\pm(\Be_{n-p+a}+\Be_{n+a})$& $[n-p]+[\overline{n-p}]$\\[7pt]
$(\Bv^+_a,\Bv^-_a)$ & $4|m_a|$ & 0& $[1]$\\[7pt]
\hline\hline     
\\[-5pt]
\end{tabular}}
\label{StateTable}
\caption{\small The perturbative states for the complex Grassmannians showing the charges under the abelian subgroup of $H$ and the representation under the $SU(n-p)$ subgroup of $H$. A charge written as $\Be_a$ corresponds to the element $E_{aa}$ of the algebra.}
}

\noindent
{\bf Example: AIII}

Consider the complex Grassmannians $SU(n+p)/S(U(n)\times U(p))$ with $n\geq p$. In this case, the eigenvalues of $\Lambda$ are
\EQ{
\big\{ im_a\big\}=\big\{\pm im_1, \ldots \pm im_p, 0^{n-p}\big\}.
}
Then, for $\Lambda$ regular, the spectrum of perturbative particles is shown in Table~\ref{StateTable}.
Taking into account that the fields in the first three rows are complex and the one in the fourth is real, the total number of perturbative states is 
 $(2n-1)p$, which matches the dimension of quotient
\EQ{
G/H=\frac{S\left(U(n)\times U(p)\right)}{S\left(U(n-p)\times U(1)^p\right)}\ .
}
The real field in the fourth row provides an example of~\eqref{RealInv}, since $\Bv_a^- = \sigma_-(\Bv_a^+)$ and $\sigma_-$ is holomorphic. In order to compare with the case of the real Grassmannians, we include the explicit expression of the charge carried by the perturbative modes associated to $(\Bv_a^+,\BOmega)$:
\EQ{
{\cal Q}=i \frac{\pi}{k} \left(\Bv_a^+ {\Bv_a^+}^\dagger + \Bv_a^- {\Bv_a^-}^\dagger - 2\BOmega \BOmega^\dagger\right)\not= {\cal Q}^\ast .
\label{ComplexCharge}
}
which confirms that they are complex.

From the mass spectrum above, it is clear that many of the states are, at this classical
level, only marginally stable. If we order the masses so that $0<m_1<m_2<\cdots$, then the excitations which are ``elementary"---in a sense which excludes marginally stable excitations---correspond to the subset
\SP{
&2(m_{a+1}-m_a)\; \longrightarrow\; p-1\;\; \text{complex fields},\\[5pt] 
& 2m_1 \; \longrightarrow\; \text{multiplet of $n-p$ complex fields}.
}
In particular, the $2(n-p)$ excitations of mass $2m_1$ transform in the $n-p$ vector (fundamental) representation of $SU(n-p)\subset   H$ and its conjugate.

\TABLE[ht]{
{\small\begin{tabular}{ccc}
\hline\hline
\\[-10pt]
Cartan      &   Mass     &      Degeneracy \\
\\[-10pt]
\hline
\\[-10pt]
AIII   &  $2|m_{a+1}-m_a|$ &  $ {\mathbb C}$ \\
                                            &  $ 2|m_1|$   &   $  [n-p]+[\overline{n-p}]$  \\[5pt]
BDI & $   2|m_{a+1}-m_a| $ & $  {\mathbb R}$\\
                                         &$        2|m_1|      $&$           [n-p]$\\[5pt]
CII  &$  2|m_{a+1}-m_a| $&$  [2_{a+1}]\times[2_a]$\\
                                        &$         2|m_1|        $&$         [2]_1\times[2(n-p)]$\\[5pt]
AI & $|m_{a+1}-m_a|$&$  {\mathbb R}$\\    [5pt]                                  
AII &$2|m_{a+1}-m_a|$ & $[2]_{a+1}\times[2]_a$\\[5pt]
DIII ($n$ even) & $2|m_{a+1}-m_a|$&$[2]_{a+1}\times[2]_a$\\ 
&$4|m_1|$&${\mathbb R}$\\[5pt]
DIII ($n$ odd)&$2|m_{a+1}-m_a|$&$[2]_{a+1}\times[2]_a$\\
&$2|m_1|$& $[2]_1+[2]_1$\\[5pt]
CI & $2|m_{a+1}-m_a|$&$  {\mathbb R}$\\
&$4|m_1|$&$  {\mathbb R}$\\[5pt]
\hline\hline
\\[-5pt]
\end{tabular}}
\label{SpecTable}
\caption{\small The Type I symmetric spaces  and the spectrum of elementary excitations with their degeneracy and $H$ representation content, in the regular case.}
}

\noindent
{\bf Example: BDI}

Consider now $SO(n+p)/SO(n)\times SO(p)$ with $n\geq p$. The eigenvalues and eigenvectors of $\Lambda$, and the action of $\sigma_-$, are identical for the complex and real Grassmannians. However, in the latter there is an additional anti-holomorphic involution $\sigma_+$ such that (eq.~\eqref{InvRealG})
\EQ{
\sigma_+(\Bv_a^\pm)=\Bv_a^\mp\ ,\qquad\sigma_+(\BOmega)=\BOmega^\ast\ .
}
For $m_b\not= -m_a$, the spectrum of perturbative states is also identical to the complex Grassmannians. But
it is easy to check that $\sigma_+$ makes all the perturbative modes real in this case. In particular, the charge of the modes associated to $(\Bv_a^+,\BOmega)$ is
\EQ{
{\cal Q}=-i \frac{\pi}{k} 2\left(\BOmega \BOmega^\dagger- \BOmega^\ast \BOmega^t\right)= {\cal Q}^\ast\ ,
}
to be compared with~\eqref{ComplexCharge}.
They transform in the vector (fundamental) representation of $SO(n-p)$, which is real.

Moreover, since $\Bv_a^+ =\sigma_+(\Bv_a^-)$ and $\sigma_+$ is anti-holomorphic, eq.~\eqref{RealInv} shows that the perturbative fields associated to $m_b=-m_a$ vanish and, hence, there are no modes corresponding to the fourth row of Table~\ref{StateTable} in this case. 
Altogether,  the total number of perturbative states is 
 $(n-1)p$, which matches the dimension of
\EQ{
G/H=\frac{SO(n)\times SO(p)}{SO(n-p)}\ .
}

\noindent
{\bf Example: AII}

In this case, the perturbative modes are associated to pairs $(\BOmega_a,\BOmega_b)$ with $a\not=b$, where $\BOmega_a$ is the generic eigenvector defined in~\eqref{EigenVAII}. The charge carried by the corresponding modes is
\SP{
{\cal Q}= i\frac{\pi}{k} \left(\BOmega_a\BOmega_a^\dagger  -\BOmega_b\BOmega_b^\dagger - J_n\BOmega_a^\ast\BOmega_a^t J_n^{-1} + J_n\BOmega_b^\ast\BOmega_b^t J_n^{-1} \right)= J_n{\cal Q}^\ast J_n^{-1} \ ,
}
which transforms in the fundamental representation of the $SU(2)\times SU(2) \subset H$ subgroup specified by $a$ and $b$, which is pseudo-real. The total number of excitations is $2n(n-1)$ which matches the dimension of $G/H = Sp(n)/SU(2)^n$.

The same analysis can be repeated for all the examples, and in Table \ref{SpecTable} we summarize the spectrum of elementary excitations and their $H$ quantum numbers.

\noindent
{\bf The $\Lambda_+\neq\Lambda_-$ generalization}

It is interesting that if we generalize the discussion temporarily to the situation with $\Lambda_+\not=\Lambda_-$ in~\eqref{www1}, some of the perturbative particles are expected to become unstable just as in the homogeneous sine-Gordon theories~\cite{FernandezPousa:1996hi,FernandezPousa:1997zb,Miramontes:1999hx,CastroAlvaredo:1999em,Dorey:2004qc}. In particular, the corresponding theories will exhibit resonance parameters. Recall that $[\Lambda_+,\Lambda_-]=0$, ensuring that they can be diagonalized in terms of the same basis of eigenvectors:
\EQ{
\Lambda_\pm \Bv_a = im_a^\pm \Bv_a\ .
}
In this case, the masses of the fundamental particles are given by 
\EQ{
\widetilde{m}_{ab} = 2\sqrt{(m_a^+- m_b^+)(m_a^--m_b^-)}\ .
}
Then, if there exist eigenvalues $m_c^+$ and $m_c^-$  such that $m_a^\pm < m_c^\pm <m_b^\pm$, then we have the bound
\EQ{
\widetilde{m}_{ab}^2 = \widetilde{m}_{ac}^2 + \widetilde{m}_{cb}^2 + 2 \widetilde{m}_{ac}\widetilde{m}_{cb} \cosh\sigma_{ab}^c\> \geq\> \left(\widetilde{m}_{ac} + \widetilde{m}_{cb}\right)^2\ ,
}
with the resonance parameter
\EQ{
\sigma_{ab}^c = \frac{1}{2} \log\left(\frac{m_a^+-m_c^+}{m_a^--m_c^-}\> \frac{m_b^--m_c^-}{m_b^+-m_c^+}\right)=-\sigma_{ba}^c\ .
}
This suggests that the particle of mass $\widetilde{m}_{ab}$, which would be marginally stable if $\Lambda_+=\Lambda_-$, will decay into the particles with masses $\widetilde{m}_{ac}$ and $\widetilde{m}_{bc}$. These fascinating theories will be investigated elsewhere, and in the rest of this work  we take $\Lambda_+=\Lambda_-$.

\section{The Integrable Hierarchy and Conserved Charges}\label{Inth}

 The fact that the SSSG theories are integrable means that they have an infinite tower of conserved charges. 
The simplest way to reveal this structure is to
write the SSSG equations in Lax form, that is as a zero curvature
condition for a connection that depends on an auxiliary
complex spectral parameter $z$:
\EQ{
{\cal L}_\mu=\partial_\mu+{\cal A}_\mu(x;z)\ ,\qquad[{\cal L}_\mu(z),{\cal L}_\nu(z)]=0\ ,
\label{zcc}
}
where 
\SP{
{\cal L}_+(z)&= \partial_++\gamma^{-1}\partial_+\gamma+\gamma^{-1}A_+\gamma-z
\Lambda
\ ,\\[5pt]
{\cal L}_-(z)&= \partial_-+A_--z^{-1}\gamma^{-1}\Lambda\gamma\ .
}
It is straightforward to check that the flatness condition yields the equation-of-motion of the SSSG theory for any value of $z$.
The proper algebraic setting for the Lax connection is the affine (loop) Lie algebra with a gradation that is fixed by the involution $\sigma_-$:
\EQ{
\hat{\mathfrak  f}= \bigoplus_{n\in\Z} \left(z^{2n} \otimes {\mathfrak g}
  +z^{2n+1} \otimes {\mathfrak p} \right)\equiv\bigoplus_{k\in
  \boldsymbol{Z}}\>\hat{\mathfrak f}_k\ ,
}
where we have defined
\SP{
\hat{\mathfrak f}_k = \begin{cases}
z^{k}\otimes \mathfrak{g}\,,      & \text{if}\;\; k=2n\,, \\
z^{k}\otimes \mathfrak{p}\,,      & \text{if}\;\; k=2n+1,,
\end{cases}
}
and $[\hat{\mathfrak f}_k,\hat{\mathfrak f}_l]\subset \hat{\mathfrak f}_{k+l}$ as a consequence of the canonical decomposition~\eqref{CanonicalDec}.
The flatness equation
provide the integrability condition for the associated linear problem
\EQ{
{\cal L}_\mu(z) \Upsilon(z)=0\ .
\label{LinProb}
}

The existence of a flat connection implies that the Wilson line 
along a curve in spacetime is independent of the curve as long as its end points are kept fixed.
Since the connection depends on an auxiliary parameter $z$, it is well known that one can use the path independence to construct an infinite set of conserved quantities. The basic idea is to consider a rectangular closed path which goes along through the points $(t,-\infty)$, $(t,\infty)$, $(t+T,\infty)$, $(t+T,-\infty)$. If the contribution from the components at $x=\pm\infty$ vanish, which requires ${\cal A}_\mu(x;z)\to 0$, then the Wilson line, or ``monodromy matrix'',
\EQ{
\text{Pexp}\,\left[-\int_{-\infty}^{\infty}dx\,{\cal A}_1(x;z)\right]=
\Upsilon(x=\infty;z)\Upsilon^{-1}(x=-\infty;z)
\label{MMatrix}
}
along the $x$ axis would be independent of time, and when expanded in powers of $z$ or $z^{-1}$ would give an infinite set of conserved quantities. However, in the present case, assuming that the fields asymptote to their vacuum values as $x\to\pm\infty$, we have 
\EQ{
{\cal A}_\pm(x;z)\longrightarrow{\cal A}_\pm^\text{vac}(z)=-z^{\pm1}\Lambda\ .
}
It follows that the monodromy defined in \eqref{MMatrix} is {\em not} conserved and is also a divergent quantity. The way to fix both problems is to define the subtracted monodromy
\EQ{
{\cal M}(z)=\lim_{x\to\infty}\Upsilon^{-1}_0(x;z)\Upsilon(x;z)\Upsilon^{-1}(-x;z)\Upsilon_0(-x;z)
\label{MMatrix2}
}
where 
\EQ{
\Upsilon_0(x;z)=\exp\big[(zx^++z^{-1}x^-)\Lambda\big]
\label{dvs}
}
is the vacuum solution of the linear problem,
and we have chosen the on-shell gauge fixing conditions $A_\mu=0$.\footnote{${\cal M}(z)$ can be defined in an equivalent way that is explicitly invariant under local gauge transformations. Since $A_\mu$ is a flat connection on-shell, we can write $A_\mu(x) = -\partial_\mu W W^{-1}$, where $W(x)=\text{Pexp}\,\left[-\int_{x_0}^{x}dx^\mu A_\mu\right]$ is a Wilson line and $x_0$ is an arbitrary reference point. Then,
\EQ{
{\cal M}(z)=\lim_{x\to\infty} W^{-1}(x)\Upsilon^{-1}_0(x;z)\Upsilon(x;z)\Upsilon^{-1}(-x;z)\Upsilon_0(-x;z) W(-x)\ .
}
Under gauge transformations, it transforms as ${\cal M}(z) \to U(x_0) {\cal M}(z)  U^{-1}(x_0)$. Therefore, it is explicitly invariant under ``local'' gauge transformations, singled out by the condition $U(x_0)=1$, and  its transformation under global (``rigid'') gauge transformations is given by~\eqref{GlobalMM}.
} 
It follows from \eqref{LinProb} that this is conserved
\EQ{
\partial_0{\cal M}(z)=0
}
and, in addition, its value for the vacuum  is ${\cal M}^\text{vac}(z)=1$.
Moreover, under global gauge transformations  it transforms as
\EQ{
{\cal M}(z)\longrightarrow U {\cal M}(z) U^{-1}.
\label{GlobalMM}
}

The expansion of the subtracted monodromy around $z=0$ and $\infty$, 
\EQ{
{\cal M}(z)=\exp\big[
q_0+q_{1}z+q_{2}z^2+\cdots\big]=\exp\big[q_{-1}/z+q_{-2}/z^2+\cdots\big]\ ,
\label{mex}
}
provide a set of conserved charges $q_s$ of Lorentz spin $s$, and we will soon show that $q_s\in{\mathfrak f}^\perp$.
Some of the charges above are given as integrals of conserved currents which are local in the Lagrangian fields, while others are non-local quantities.\footnote{
The calculation of the conserved charges carried by the soliton solutions of 1 + 1 dimensional integrable field theories in terms of their asymptotic spatial behaviour have also been addressed in~\cite{Olive:1992iu,Miramontes:1998hu,Ferreira:2007pr}.}
In particular, we shall argue that the null components of the conserved 2-momentum of a configuration are given by
\EQ{
p_\pm=\mp\frac k{2\pi}\Tr(\Lambda q_{\mp1})\ .
\label{tmn}
}

The form of the conserved currents can be deduced using the Drinfeld-Sokolov procedure~\cite{DeGroot:1991ca}. We start by considering the currents of positive spin and introduce\footnote{In the following we will often use the notation
$
\hat{\mathfrak f}_{<0}=\bigoplus_{k<0}\hat{\mathfrak f}_k$, $\hat{\mathfrak f}_{\geq0}=\bigoplus_{k\geq0}\hat{\mathfrak f}_k$,
{\it etc\/}.
}
\EQ{
\Phi=\exp\,y(z)\>,\qquad y(z)=\sum_{s\geq1}z^{-s}y_{-s} \in \hat{\mathfrak f}_{<0}
}
and solve
\EQ{
\Phi^{-1} {\cal L}_+(z) \Phi=  \partial_+  - z\Lambda+ h_+(z)\ ,\qquad
h_+(z)=\sum_{s\geq0} h_{-s,+} z^{-s}\in\hat{\mathfrak f}^\perp_{\leq0} \ .
\label{DS}}
Correspondingly,
\EQ{
\Phi^{-1} {\cal L}_-(z)\Phi = \partial_-  +h_-(z)\ , \qquad h_-(z)\in \hat{\mathfrak f}_{\leq0}
\label{DS2}
}
and the zero curvature condition~\eqref{zcc} implies
\EQ{
\bigl[\partial_+ -z\Lambda+ h_+(z),  \partial_-  + h_-(z)\bigr]=0\>,
\label{Zero2}
}
from which it follows
\EQ{
h_-(z)=\sum_{s\geq0}h_{-s,-}z^{-s}\in \hat{\mathfrak f}^\perp_{\leq0}\ .
}
The zero-curvature condition \eqref{Zero2} proves directly that the projection of $h_\pm(z)$ onto ${\mathfrak z}(\Lambda)$, the centre of $\hat{\mathfrak f}^\perp$, lead to conserved currents
\EQ{
J^\mu(z)=\epsilon^{\mu\nu}h_\nu(z)\Big|_{{\mathfrak z}(\Lambda)}\ .
}
Since ${\mathfrak z}(\Lambda)$ always contains
the infinite set of elements $z^{2n+1}\Lambda$,
as well as any abelian factors of ${\mathfrak h}$ times $z^{2n}$, 
there are an infinite number of local conserved charges.

In~\cite{Miramontes:1998hu}, it was emphasized that the choice of $\Phi$ is not unique. It is defined modulo the transformations $\Phi\rightarrow \Phi\eta$, with
\EQ{
\eta \in \exp \,\hat{\mathfrak f}^\perp_{<0}\>,
}
which does not change the form of~\eqref{DS}, but does change the value of $h_\pm(z)$. This freedom can be used to ensure that $\Phi$ and $h_\pm(z)$ are local functions of the component fields by simply enforcing the condition
\EQ{
y(z)\in  \hat{\mathfrak f}^\parallel_{<0}\>,
\label{LocalCond}
}
which will always be assumed in the following.\footnote{More precisely, $\Phi$ and $h_+(z)$ are local functions of the combination of fields ${\cal L}_+(z)+ z\Lambda=\gamma^{-1}\partial_+\gamma +\gamma^{-1}A_+\gamma$.}

In a similar way, a second set of conserved densities with negative spin can be constructed starting from
\EQ{
\gamma{\cal L}_-(z)\gamma^{-1}&= \partial_- -\partial_-\gamma\gamma^{-1}+ \gamma A_-\gamma^{-1} - z^{-1}\Lambda\ ,\\
\gamma{\cal L}_+(z)\gamma^{-1}&=\partial_++A_+-z\gamma\Lambda\gamma^{-1}
\label{DSMinus}
}
instead of ${\cal L}_\pm$, with 
\EQ{
\Phi\rightarrow \tilde\Phi\in \exp \hat{\mathfrak f}^\parallel_{>0}\ ,\qquad
h_\mu(z)\rightarrow \tilde h_\mu(z)\in \hat{\mathfrak f}^\perp_{\geq0}\ .
}
and
\EQ{
\tilde h_\mu(z)=\sum_{s\geq 0} h_{s,\mu}z^s\ .
}
Both constructions, and in particular the two quantities $h_\mu(z)$ and $\tilde h_\mu(z)$, are trivially related by means of the replacements
\EQ{
z\rightarrow z^{-1},\quad
\partial_+\rightarrow \partial_-,\quad
\gamma\rightarrow \gamma^{-1},\quad
A_\pm\rightarrow A_\mp\ .
\label{Parity}
}

To illustrate how concrete the construction 
is, we include the explicit expressions for the densities of spin~1 and~2. They can be found by looking at the first components of~\eqref{DS}, which read
\AL{
&
h_{0,+}- [y_{-1},\Lambda]=\gamma^{-1}\partial_+\gamma + \gamma^{-1}A_+\gamma\equiv a_+ \label{Recurr1}\ ,\\[5pt]
&
h_{-1,+}-[y_{-2},\Lambda] =\partial_+ y_{-1} -[y_{-1},a_+]-\tfrac{1}{2}[y_{-1},[y_{-1},\Lambda]]\ .
\label{Recurr2}
}

Projecting \eqref{Recurr1} onto ${\mathfrak f}^\perp$ and using 
~\eqref{Orthogonal} and~\eqref{LocalCond} as well as the constraint \eqref{gco2}, one gets
\EQ{
h_{0,+}=a_+^\perp=\Big(\gamma^{-1}\partial_+\gamma + \gamma^{-1}A_+\gamma\Big)^\perp\equiv A_+\>.
}
In turn,~\eqref{DS2} gives
\EQ{
h_{0,-}= A_-\ .
 }
Therefore, the 0-grade component of~\eqref{Zero2}
is just the flatness condition \eqref{flat}. Projecting now \eqref{Recurr1} onto ${\mathfrak f}^\parallel$ implicitly determines $y_{-1}$:
\EQ{
[y_{-1},\Lambda]=-a_+^\parallel\ .
}

The densities of spin~2 provide the components of the stress-energy tensor, which are the projections of $h_{\pm1,\pm}$  along $\Lambda$. Using~\eqref{Recurr2},
\SP{
\frac\kappa{2\pi}{\rm Tr}\bigl(\Lambda h_{-1,+}\bigr)&=-\frac\kappa{2\pi}{\rm Tr}\Big[\Lambda\Big([y_{-1},a_+]+\frac{1}{2}[y_{-1},[y_{-1},\Lambda]]\Big)\Big]\\[5pt]
&=-\frac{\kappa}{4\pi}{\rm Tr}\big( a_+^2-(a^\perp_+)^2\big)\equiv T_{++}\ .
\label{DensTa}
}
Correspondingly,
\EQ{
\frac\kappa{2\pi}{\rm Tr}\bigl(\Lambda h_{-1,-}\bigr)=-\frac\kappa{2\pi}{\rm Tr}\Bigl(\Lambda\gamma^{-1}\Lambda\gamma \Bigr)\equiv -T_{-+}\>,
\label{DensTb}
}
and~\eqref{Zero2} leads to
\EQ{
\partial_+T_{-+}+
\partial_-T_{++}=0\ .
}
The component $T_{--}$ is obtained is a similar fashion starting from~\eqref{DSMinus}, or using~\eqref{Parity}.
Then, the complete set of components of the energy-momentum tensor can be written as
\AL{
T_{++}&=-\frac{\kappa}{4\pi}\Tr\Big[\bigl(\partial_+\gamma\gamma^{-1}+ A_+\bigr)^2-A_+^2\Bigr]
\label{Tplusplus}
\\[5pt]
T_{--}&=-\frac{\kappa}{4\pi}\Tr\Big[\bigl(\gamma^{-1}\partial_-\gamma- A_-\bigr)^2 -A_-^2\Bigr]
\label{Tminusminus}
\\[5pt]
T_{-+}&=T_{-+}=\frac\kappa{2\pi}{\rm Tr}\Big[\Lambda\gamma^{-1}\Lambda\gamma
\Big]\>,
\label{ener}
}
which are explicitly invariant under both local and global gauge transformations.
This process can be continued and at order $z^n$, $y_{n+1}$ is determined algebraically in terms of $a_+$ and $y_s$ with $s\leq n$. This proves our assertion that $y(z)$ is a local function of of the fields.

In the following, we will choose the on-shell gauge fixing conditions $A_\mu=0$, which are enabled by the flatness condition~\eqref{flat} and imply
\EQ{
h_\mu \in \hat{\mathfrak f}^\perp_{<0}, \qquad
\tilde h_\mu \in \hat{\mathfrak f}^\perp_{>0}\ .
}
Then, in order to deduce the relationship between the conserved densities and the subtracted monodromy~\eqref{MMatrix2}, we solve the zero curvature condition~\eqref{Zero2} as follows
\EQ{
h_+(z)= \Omega\partial_+ \Omega^{-1}\>,\qquad
h_-(z)= -z^{-1}\Lambda+ \Omega\partial_- \Omega^{-1}\>,
\qquad
\Omega\in \exp \hat{\mathfrak f}^\perp_{<0}\ .
}
This leads to
\EQ{
\chi^{-1}{\cal L}_\pm(z)\chi=\partial_\pm -z^{\pm1} \Lambda\>,\qquad
\chi = \Phi\Omega \in \exp \hat{\mathfrak f}_{<0}\ .
\label{MiniLax1}
}
In other words, $\chi=\chi(z)$ is a formal series in $z^{-1}$ taking values in $F$ normalized such that $\chi=1$ at $z=\infty$. This provides the following expression for the solution to the associated linear problem~\eqref{LinProb}: 
\EQ{
\Upsilon(z)= \chi(z) \Upsilon_0(z) g_+(z),
\label{LPsol1}
}
where $\Upsilon_0(z)$ is the vacuum solution defined in \eqref{dvs} and
$g_+(z)$ is a constant element of the loop group.
In a completely analogous fashion, starting from $\gamma{\cal L}_-(z)\gamma^{-1}$ instead of ${\cal L}_+(z)$ we get
\EQ{
\tilde\chi^{-1}\gamma{\cal L}_\pm(z)\gamma^{-1}\tilde\chi=\partial_\pm -z^{\pm1} \Lambda\>,\qquad
\tilde\chi=  \tilde\Phi\tilde\Omega\in \exp \hat{\mathfrak f}_{>0}\>,
\label{MiniLax2}
}
where
\EQ{
\tilde h_+(z)= -z\Lambda +\tilde \Omega\partial_+ \tilde\Omega^{-1}\>,\qquad
\tilde h_-(z)= \tilde\Omega\partial_-\tilde\Omega^{-1}\>.
}
In this case, $\tilde\chi=\tilde\chi(z)$ is a formal series in $z$ normalized such that $\tilde\chi=1$ at $z=0$. Then,~\eqref{MiniLax2} provides another expression for the solution to the associated linear problem:
\EQ{
\Upsilon(z)= \gamma^{-1} \tilde\chi(z) \Upsilon_0(z) g_-(z)\ ,
\label{LPsol2}
}
where $g_-(z)$ is another constant element of the loop group. Equating~\eqref{LPsol1} and~\eqref{LPsol2} gives rise to the factorization (Riemann-Hilbert) problem\footnote{Notice that the normalization of $\chi(z)$ and $\tilde\chi(z)$ at $z=\infty$ and $z=0$, respectively, is fixed by our choice of gauge fixing conditions.}
\EQ{
\Upsilon_0(z) g_-(z)g_+(z)^{-1} \Upsilon_0^{-1}(z)=\tilde\chi(z)^{-1}\gamma\chi(z)\>.
\label{RHp}
}

Eqs.~\eqref{LPsol1} and~\eqref{LPsol2} lead to two alternative expressions for the subtracted monodromy:
\EQ{
&{\cal M}(z)\\
&=\lim_{x\to\infty}\Upsilon^{-1}_0(x;z)\chi(x;z)
\Upsilon_0(x;z)\Upsilon^{-1}_0(-x;z)\chi^{-1}(-x;z)
\Upsilon_0(-x;z)\ ,\\[5pt]
&=\lim_{x\to\infty}\Upsilon^{-1}_0(x;z)\gamma^{-1}(x)\tilde\chi(x;z)
\Upsilon_0(x;z)\Upsilon^{-1}_0(-x;z)\tilde\chi^{-1}(-x;z)\gamma(x)
\Upsilon_0(-x;z)
\ .
}
Then, assuming that the currents $\gamma^{-1}\partial_+\gamma$ and $\partial_-\gamma\gamma^{-1}$  
fall off sufficiently fast at infinity,  and since $\Phi$ and  $\tilde\Phi$ depend locally on them, we have
\EQ{
\lim_{x\to\pm\infty}\Phi(x;z)=1\ ,\qquad
\lim_{x\to\pm\infty}\tilde\Phi(x;z)=1
}
and so
\EQ{
\lim_{x\to\pm\infty}\chi(x;z)=\lim_{x\to\pm\infty}\Omega(x;z)\ ,\qquad\lim_{x\to\pm\infty}\tilde\chi(x;z)=\lim_{x\to\pm\infty}\tilde\Omega(x;z)\ .
}
In addition, since $\Omega,\tilde\Omega\in\exp{\mathfrak f}^\perp$ and $\gamma(\pm\infty)\in H=\exp{\mathfrak h}$, this means that $\chi(\pm\infty;z)$ and, hence, $\tilde\chi(\pm\infty;z)\in\exp{\mathfrak f}^\perp$ commute with $\Upsilon_0(x;z)$. So the subtracted monodromy 
is finally given by the two expressions:
\AL{
{\cal M}(z)&=\chi(\infty;z)\chi^{-1}(-\infty;z)=
\text{Pexp}\,\left[-\int_{-\infty}^{+\infty}dx\, \big(h_1(z)-z^{-1}\Lambda\big)\right]
\label{ExpInftyB}\\[7pt]
&=\gamma^{-1}(\infty)\tilde\chi(\infty;z)
\tilde\chi^{-1}(-\infty;z)\gamma(-\infty)\notag\\[5pt]
&=\gamma^{-1}(\infty)\> \text{Pexp}\,\left[-\int_{-\infty}^{+\infty}dx\, \big(\tilde h_1(z)+z\Lambda\big)\right]\gamma(-\infty)\>.\label{ExpZeroB}
}

Expanding \eqref{ExpZeroB} around $z=0$ as in \eqref{mex} gives directly the kink charge of a configuration as in \eqref{chgs2}.\footnote{We choose to call it kink charge rather than topological charge since it is not quantized at the classical level. However, in the quantum theory it will, indeed, be quantized.} Using~\eqref{ExpInftyB}, \eqref{ExpZeroB}, \eqref{DensTa} and~\eqref{DensTb}, it is easy to check that
\EQ{
\mp(\kappa/2\pi)\Tr(\Lambda q_{\mp1})=\int_{-\infty}^{+\infty} dx \>\left[T_{0\pm} -T_{0\pm}^{\text{vac}} \right]=p_\pm.
}
The subtraction  in the above corresponds to the fact that the vacuum configuration $\gamma=1$ has $T_{+-}^\text{vac}=(\kappa/2\pi)\Tr(\Lambda^2)$, from \eqref{ener}, while ${\cal M}^\text{vac}(z)=1$. The physical momentum of a configuration is then defined relative to the vacuum.

\section{The Solitons}
\label{Dressing}

Soliton solutions have been constructed in the SSSG theories in
\cite{Hollowood:2009tw,Hollowood:2009sc} by means of the dressing method~\cite{Zakharov:1973pp}. However, in those works the Pohlmeyer
reduced $F/G$ sigma model played a prominent role. In the present
context, we set up the dressing method directly in the SSSG theories. 
Although the solutions are identical to
those found earlier, the formalism intrinsic to the SSSG theories is somewhat simpler and Lorentz symmetry is manifest.

The dressing method naturally produces solutions to the
equations-of-motion taking values in the complex group $GL(N,{\mathbb C})$. It
is therefore necessary to impose reality conditions on the basic solution so that  the Lax connection is valued in the affine algebra $\hat{\mathfrak f}$. To achieve this, and since $F\subset SU(N)$, we must first impose anti-hermiticity:
\EQ{
{\cal L}_\mu(z)^\dagger=-{\cal L}_\mu(z^*)
\label{cond1}
}
and then conditions for each of the involutions associated to each case:
\EQ{
\sigma_+\big({\cal L}_\mu(z)\big)= {\cal L}_\mu(\widetilde z)\ ,\qquad
\sigma_-\big({\cal L}_\mu(z)\big)= {\cal L}_\mu(-\widetilde z)\ ,
\label{cond2}
}
where $\widetilde z= z$, or~$z^\ast$, if $\sigma_\pm$ are holomorphic, or anti-holomorphic, respectively. For the consistency of the associated linear problem, these conditions
have to be lifted to the group $F$ itself and imposed as 
\EQ{
\Upsilon(z)^\dagger=\Upsilon(z^*)^{-1}\ ,\qquad \sigma_+\big(\Upsilon(z)\big)=\Upsilon(\widetilde z)\ , \qquad
\sigma_-\big(\Upsilon(z)\big)= \Upsilon(-\widetilde z)\ .
\label {cond3}
}
Note that these constraints are trivially satisfied by the vacuum solution~\eqref{dvs}.

Solitons are special solutions 
for which $g_+(z)=g_-(z)=1$ in the Riemann-Hilbert problem \eqref{RHp}~\cite{Babelon}. Then,~\eqref{LPsol1} and~\eqref{LPsol2} imply that the solution of the linear problem can be written in two equivalent ways
\EQ{
\Upsilon(x;z)=\chi(x;z)\Upsilon_0(x;z)=\gamma^{-1}\tilde\chi(x;z)\Upsilon_0(x;z)\ .
\label{dres}
}
In the context of solitons, the first of these
is known as the ``dressing transformation" for the obvious reason that it ``generates" the soliton solutions from the vacuum, and $\chi(z)\equiv\chi(x;z)$ is called the dressing factor.
Eq.~\eqref{dres} shows that it can be expanded around both $z=0$ and $z=\infty$. 
The soliton solutions have $A_\mu=0$, and 
the associated linear problem~\eqref{LinProb} gives rise to the two equations
\AL{
\partial_+\chi(z) \chi(z)^{-1} + z\chi(z) \Lambda\chi(z)^{-1}& = -
\gamma^{-1}\partial_+\gamma+z\Lambda\ ,
\label{Eq1}\\[5pt]
\partial_-\chi(z) \chi(z)^{-1} + z^{-1}\chi(z) \Lambda\chi(z)^{-1}& = z^{-1}\gamma^{-1}\Lambda\gamma\ .
\label{Eq2}
}
Identifying the residues of the both sides of~\eqref{Eq1} and ~\eqref{Eq2} at $z=\infty$ and $z=0$, respectively, we find
\EQ{
\chi(\infty)=1, \qquad
\chi(0)=\gamma^{-1}\ .
\label{NormalDF}
}
The relations above do not necessarily ensure that $\det\>\gamma=1$ and so may involve a compensating scalar factor that is a power of $\det\>\chi(0)$. However, as we shall be interested in calculating the charges of the soliton, this complication is irrelevant and we suppress it. If we define the series around $\infty$ and $0$
\EQ{
\chi(z)=1+W_+z^{-1}+{\cal O}(z^{-2})\ ,\qquad
\chi(z)=\gamma^{-1}\big(1+W_-z+{\cal O}(z^2)\big) \ ,
}
then we have
\EQ{
\gamma^{\mp1}\partial_\pm\gamma^{\pm1}=[\Lambda,W_\pm]\ .
}
Consequently, the soliton satisfies the on-shell constraints \eqref{gcond3}.\footnote{We remark once more that the normalization conditions $\chi(\infty)=\tilde\chi(0)=1$ are associated to the gauge fixing conditions $A_\mu=0$.}

The dressing method then proceeds by taking an ansatz for the dressing factor which takes the form of a sum over a finite set of simple poles
\EQ{
\chi(z)=1+\frac{Q_i}{z-\xi_i}\ ,\qquad
\chi(z)^{-1}=1+\frac{R_i}{z-\mu_i}\ .
}
Since the right-hand side of  \eqref{Eq1} and \eqref{Eq2} is regular at 
$z=\xi_i$ and $\mu_i$ the residues of the left-hand side must vanish, giving
\SP{
\left(\xi_i^{\mp1} \partial_\pm Q_i + Q_i
  \Lambda\right)\Big(1+\frac{R_j}{\xi_i-\mu_j}\Big) &=0\ ,\\[5pt]
\Big(1+\frac{Q_j}{\mu_i-\xi_j}\Big)\left(-\mu_i^{\mp1}\partial_\pm
  R_i + \Lambda R_i\right)&=0\ .
}
These equations are very similar to those considered in~\cite{Hollowood:2009tw}. The key  
to solving them is to propose that the residues have rank one~\cite{Harnad:1983we}:
\EQ{
Q_i = \BX_i \BF_i^\dagger \quad\text{and}\quad
R_i = \BH_i \BK_i^\dagger\ .
}
The solution reads
\EQ{
\BF_i=\big(\Upsilon_0(\xi_i)^\dagger\big)^{-1}\Bvarpi_i\ ,\qquad
\BH_i=\Upsilon_0(\mu_i)\Bpi_i\ ,
}
for constant complex $N$-vectors $\Bvarpi_i$ and $\Bpi_i$ along with
\EQ{
\BX_i\Gamma_{ij}=\BH_j\ ,\qquad \BK_i(\Gamma^\dagger)_{ij}=-\BF_j\ ,
}
where the matrix
\EQ{
\Gamma_{ij}=\frac{\BF_i^*\cdot \BH_j}{\xi_i-\mu_j}\ .
}

The conditions \eqref{cond1} and \eqref{cond2} must now be imposed on
the raw solution and this determines the number of poles and
the relations between the constant vectors $\Bvarpi_i$ and $\Bpi_i$ required to produce a single irreducible soliton.
The unitarity condition \eqref{cond1} requires
\EQ{
\mu_i=\xi_i^\ast\ ,\qquad\qquad\BH_i=\BF_i=\Upsilon_0(\xi_i^\ast)\Bpi_i\ ,\qquad\BK_i=\BX_i\ .
}
which imply
\EQ{
\Bpi_i=\Bvarpi_i\ .
}
The conditions \eqref{cond2}
mean that the set of poles $\{\xi_i\}$ must
be invariant under $\xi\to \tilde\xi$ (for $\sigma_+$), or
$\xi\to -\tilde\xi$ (for $\sigma_-$). Consequently, applying the
constraints means that the poles must come in sets of $n_0$ elements:
\EQ{
\xi_i=\sigma_i(\xi)\ ,\qquad\Bvarpi_i=\sigma_i(\Bvarpi)\ .
}
where we have ordered the elements of $\II$ as $\{1,\sigma_-\}$ and $\{1,\sigma_+,\sigma_-,\sigma_-\sigma_+\}$, for $n_0=2$ and 4, respectively. The results are written in 
Table~\ref{RankTable}. The Table also shows the additional constraints of the form
\EQ{
\Bvarpi_i^\ast\cdot  \Bvarpi_j=0
\label{extc}
}
which must be imposed on the dressing data
whenever $\xi_j=\xi_i^*$, since
this ensures that the element $\Gamma_{ij}=0$ rather than being
na\"\i vely divergent. Notice, however, that this constraint is automatically satisfied if $\Bvarpi_i=J_n\Bvarpi_j^*$ due to the anti-symmetry of $J_n$ (this occurs for cases CI and CII).

\TABLE[ht]{
{\small\begin{tabular}{cccc}
\hline\hline
\\[-10pt]
Cartan &   Dressing Poles & Dressing vectors & Constraints
\\
\\[-10pt]
\hline
\\[-10pt]
AIII &   $\{\xi,-\xi,\}$ & $\{\Bvarpi,I_{np}\Bvarpi\}$  & \\[5pt]
BDI &    $\{\xi,\xi^*,-\xi,-\xi^*\}$ & $\{\Bvarpi,\Bvarpi^*,I_{np}\Bvarpi,I_{np}\Bvarpi^*\}$ & $\Bvarpi\cdot \Bvarpi=0$ \\[5pt]
CII &     $\{\xi,\xi^*,-\xi,-\xi^*\}$ & $\{\Bvarpi,J_{n+p}\Bvarpi^*,K_{np}\Bvarpi,J_{n+p}K_{np}\Bvarpi^*\}$ & \\[5pt]
AI &    $\{\xi,-\xi^*\}$ & $\{\Bvarpi,\Bvarpi^*\}$ & \\[5pt]
AII &    $\{\xi,-\xi^*\}$ & $\{\Bvarpi,J_n\Bvarpi^*\}$ & \\[5pt]
DIII &    $\{\xi,\xi^*,-\xi,-\xi^*\}$ & $\{\Bvarpi,\Bvarpi^*,J_n\Bvarpi,J_n\Bvarpi^*\}$ & $\Bvarpi\cdot\Bvarpi=0$ \\[5pt]
CI &   $\{\xi,\xi^*,-\xi,-\xi^*\}$&$\{\Bvarpi,J_n\Bvarpi^*,J_n\Bvarpi,\Bvarpi^*\}$ & \\
[5pt]
\hline\hline
\\[-5pt]
\end{tabular}}
\label{RankTable}
\caption{\small The Type I symmetric spaces and the dressing data of their solitons.}
}

\noindent
{\bf Collective Coordinates}

The data $(\Bvarpi,\xi)$  are the parameters associated to
the solution. At the moment, we cannot call them ``collective
coordinates'' in the usual sense because these are defined as the
parameters associated to the most general solution of a given energy. In fact it will
transpire that $\xi$ is not a collective
coordinate but rather determine the mass and velocity of a
solution, whereas $\Bvarpi$ {\em is} a genuine
internal collective coordinate once the elementary solutions are
identified.

A Lorentz transformation $x^\pm\rightarrow e^{\mp\vartheta} x^\pm$ is equivalent to the rescaling $z\rightarrow e^{\vartheta} z$, and this has the
effect of scaling the positions of the pole by $e^{-\vartheta}$; namely $\xi\to e^{-\vartheta}
\xi$. Hence, the rapidity of the soliton is given by $e^{
\vartheta}=|\xi|^{-1}$, and it is worth noticing that $|\xi|=|\sigma(\xi)|$ for all $\sigma\in{\mathscr I}$.
This can be easily checked by considering the vector\EQ{
\BF\equiv\BF_1=  \Upsilon_0(\xi^\ast) \Bvarpi=\exp\left[(2t' \cos q + 2ix' \sin q)\Lambda\right]\Bvarpi\ , \qquad \xi =e^{-\vartheta -iq}\ ,
}
where $\xi\equiv\xi_1$, $\Bvarpi\equiv\Bvarpi_1$ and, moreover,
\EQ{
t' = t\cosh\vartheta -x\sinh\vartheta\ ,\qquad
x' = x\cosh\vartheta -t\sinh\vartheta
}
are the Lorentz boosted coordinates corresponding to the velocity $v=\tanh \vartheta$.

In order to identify the irreducible solitons, we can expand $\Bvarpi$ in terms of the 
eigenvectors of $\Lambda$ in the form
\EQ{
\Bvarpi=\sum_ay_a\Bv_a\ .
\label{wve}
}
Then,
\EQ{
\BF=\sum_a\exp\big(2im_at\cos q-2m_ax\sin q\big)y_a\Bv_a\ .
}
Such a solution will consists of a number of constituents whose relative positions are controlled by the ratios $y_a/y_b$.
In fact, notice that constant shifts of the solution in
space and time act on the coefficients $y_a$ via
\EQ{
y_a\longrightarrow e^{2m_a \left(i \delta t \cos q - \delta x \sin q\right)}
y_a\ .
\label{com}
}
Notice also that the soliton solution is invariant under the scaling $\Bvarpi\to \lambda\Bvarpi$. This shows that only the ratios $y_b/y_a$ are physical and, if $m_a\not=m_b$, they control the relative positions of the constituents. Therefore, in order to have a potential elementary soliton, we must restrict to cases with only two eigenvectors with different eigenvalues:
\EQ{
\Bvarpi = y_a \Bv_a + y_b \Bv_b\ , \qquad m_a\not=m_b\ .
\label{der}
}
Then, the ratio $y_a/y_b$ fixes the spacetime position of the soliton and, up to constant shifts in~$t$ and~$x$ and using the overall scaling symmetry, we can take $y_a=y_b=1$. 

If the eigenvalues of $\Lambda$ are degenerate then the eigenvectors lie in degenerate subspaces and the soliton carries internal degrees-of-freedom corresponding to the choice of the direction in the subspace. This choice is precisely correlated with the fact that degeneracies of $\Lambda$ correspond to particular subgroups of $H$ which act on $\Bvarpi$ as 
\EQ{
\Bvarpi\longrightarrow U\Bvarpi\ ,\qquad U\in H\ .
}
This action can also be thought of as the action of a the global gauge transformations:
\EQ{
\gamma\longrightarrow U\gamma U^{-1}\ ,\qquad U\in H\ .
\label{tact}
}

Let us go back to the soliton specified by the data $\Bvarpi = \Bv_a+ \Bv_b$ and $\xi=e^{-\vartheta -iq}$, and assume that $m_a<m_{b}$. If we choose $q$ such that $\sin q>0$, then in the asymptotic regimes $x\to+\infty$ and $x\to-\infty$ we can replace $\Bvarpi$ with $\Bv_{a}$ and $\Bv_{b}$, respectively.\footnote{If $\sin q<0$, then the asymptotic regions are interchanged.}
In fact, exploiting the scaling symmetry of the solution, as $x\to\pm\infty$ we can effectively replace $\BF$ with $\Bv_{a}$ and $\Bv_{b}$. It is then a simple matter to calculate the dressing factor $\chi(z)$ in the asymptotic regimes and extract the subtracted monodromy via \eqref{ExpInftyB}:
\EQ{
 {\cal M}(z)=\Big(1+\frac{\sigma_i(\Bv_{a})[\Gamma^{(a)}]^{-1}_{ij}\sigma_j(\Bv_{a})^\dagger}
{z-\sigma_j(\xi)}\Big)
\Big(1-\frac{\sigma_k(\Bv_{b})[\Gamma^{(b)}]^{-1}_{kl}\sigma_l(\Bv_{b})^\dagger}
{z-\sigma_k(\xi)^*}\Big)\ ,
}
where we have defined
\EQ{
\Gamma^{(a)}_{ij}=\Gamma_{ij}\Big|_{\BF_i\to \sigma_i(\Bv_a)}=
\frac{\sigma_i(\Bv_a)^*\cdot \sigma_j(\Bv_a)}{\sigma_i(\xi)-\sigma_j(\xi)^*}\ .
}
For most cases, (with a choice of basis for the degenerate cases)
\EQ{
\sigma_i(\Bv_a)^*\cdot \sigma_j(\Bv_a)=\delta_{ij}\ .
\label{frr}
}
The exceptions are:

(i) For AI for which the eigenvectors are real and so $\Bv_a=\sigma_+(\Bv_a)$.

(ii) For the first 3 cases, $\Lambda$ has null eigenvectors $\BOmega$ for which 
\EQ{
&I_{np}\BOmega=\BOmega\ , \qquad \text{AIII}\quad \text{and}\quad
\text{BDI}\ ,\\[5pt]
&K_{np}\BOmega=\BOmega\ , \qquad \text{CII}\ ,
}
and so $\BOmega^\ast=\sigma_+(\BOmega)$ and  $\BOmega=\sigma_-(\BOmega)$, respectively.

In cases where \eqref{frr} holds, $\Gamma_{ij}^{(a)}$ is diagonal and so we can immediately write down the expression for the log of the subtracted monodromy
\EQ{
\log {\cal M}(z)=\text{sign}(\sin q)\sum_{i=1}^{n_0}
\log\left[\frac{z-\sigma_i(\xi)^*}{z-\sigma_i(\xi)}\right]\sigma_i\left(\Bv_{a}\Bv_{a}^\dagger
-\Bv_{b}\Bv_{b}^\dagger\right)\ ,
\label{smo}
}
where we have written the result so that it is valid for either sign of $\sin q$.
Notice that $\sigma(\Bv_a)$, $\sigma\in\II$, are eigenvectors of $\Lambda$ and so the monodromy manifestly commutes with $\Lambda$. One can then verify by direct computation that the formula also covers the exceptional cases (i) and (ii) above, for which $\Gamma_{ij}^{(a)}$ is not diagonal and so it is completely general. 

From \eqref{smo}, we find the kink charge
\EQ{
q_0=+2\,\text{sign}(\sin q)\,q\,\sum_{\sigma\in\II}
\sigma\left(i\Bv_{a}\Bv_{a}^\dagger
-i\Bv_{b}\Bv_{b}^\dagger\right)\ .
\label{topc}
}
From \eqref{tmn}, the 2-momenta are
\EQ{
p_\pm&=\mp \frac\kappa{2\pi}\Tr(\Lambda q_{\mp1})\\
&=\mp\frac{ i\kappa}{2\pi}\text{sign}(\sin q)\sum_{\sigma\in\II}\big(\sigma(\xi)^{\pm1}-\sigma(\xi^*)^{\pm1}\big)\big(
\sigma(m_a)-\sigma(m_{b})\big)\\
&=\frac{2k}\pi\left|(m_{b}-m_a)\sin q\right|e^{\mp\vartheta}\ ,
}
where we used \eqref{kapk}; namely, $n_0\kappa = 2k$. So
the mass of the soliton is
\EQ{
M=\frac{4k}\pi\left|(m_{b}-m_a)\sin q\right|\ .
\label{mfr}
}
Notice that the monodromy is invariant under $q\to-q$. This means that although the vacua at $x=\pm\infty$ change the solution represents the same physical soliton with the same set of charges. Consequently, the range of $q$ can be restricted to $0<q<\pi$.

\noindent
{\bf Comparison with the perturbative excitations}

It is remarkable that the spectrum of solitons matches very closely with the perturbative quanta. Both types of excitation are associated to the same data in the form of two eigenvectors $(\Bv_a,\Bv_b)$ of $\Lambda$ with $m_a\neq m_b$. The masses and kink/$  H$ charges are also closely related:
\EQ{
M_\text{soliton}=\frac{2k}\pi|\sin q|\cdot M_\text{quantum}\ ,
}
and
\EQ{
q_0=-\frac{2kq}{\pi}\,\big(\phi(x=\infty)-\phi(x=-\infty)\big)\ .
}
In fact, the agreement is perfect in the semi-classical limit $k\gg1$ if $q=\pi/(2k)$. We will see how this special value of $q$ arises later.

We can firm up the relation between the solitons and perturbative quanta by taking the  limit $q\to0$ of the soliton solution (with $x$ finite). Then, writing $\gamma\simeq 1+\phi+\cdots$, we find to leading order (assuming $m_a<m_b$)
\EQ{
\phi^\parallel&\thicksim -q\sum_{\sigma\in\II}\sigma\Big(
i\>e^{-ip\cdot x}\>\Bv_a\Bv_b^{\dagger}
+ie^{+ip\cdot x}\>\Bv_b\Bv_a^{\dagger}\Big)
\label{ParSol}
}
with $p\cdot x = 2|m_b-m_a|\big(t\cosh\vartheta - x\sinh\vartheta)$
which, compared with~\eqref{PertField} and~\eqref{modes2}, takes the form of an on-shell particle mode with mass $M=2|m_b-m_a|$ and rapidity $\vartheta$.
Correspondingly,
\EQ{
\phi^\perp \thicksim
-2q\> \sum_{\sigma\in\II}
\sigma\left(i\> \frac{e^{4m_b x' q}\>\Bv_a\Bv_a^\dagger + e^{4m_a x' q}\>\Bv_b\Bv_b^\dagger}{e^{4m_ a x' q } +e^{4m_ b x' q }}\right)
\label{PerpSol}
}
with $x' = x\cosh\vartheta -t\sinh\vartheta$,
which exhibits the kink-like behaviour
\EQ{
\phi^\perp(\infty)-\phi^\perp(-\infty)= -2 q\,\sum_{\sigma\in\II}
\sigma\left(i\Bv_{a}\Bv_{a}^\dagger
-i\Bv_{b}\Bv_{b}^\dagger\right) =-q_0.
}
It is worth noticing that, in contrast to $\phi^\parallel$, the field $\phi^\perp$ is actually a kink and the limits $q\to0$ and $x\to\pm\infty$ do not commute. This is the reason why we have not fully approximated~\eqref{PerpSol} to leading order in $q$. However, it is easy to check that $\partial_\pm\phi^\perp$ is not kink-like, and that~\eqref{ParSol} and~\eqref{PerpSol} do satisfy the constraint~\eqref{wer} at leading order in $q$.

The soliton solution we have constructed above is a candidate for an elementary soliton. However, there is no guarantee that it cannot be split in more elementary constituents.
Recall that the value of the conserved quantities carried by a solution with  $\Bvarpi$ as in \eqref{wve} is fixed by the $x\to\pm\infty$ behaviour of $\BF$, which depend only on the highest and the lowest eigenvalues of $\Lambda$ for which $y_a\not=0$. Consequently, if for the soliton solutions with $\Bvarpi=\Bv_a+\Bv_b$ there exists an eigenvalue $m_c$ with $m_b>m_c>m_a$ (assuming that $m_b>m_a$), the solution  is a special case of the more general one with 
\EQ{
\Bvarpi = \Bv_a + y_c  \Bv_c+\Bv_b\ ,
}
which has---at least---two constituents with a relative position determined by $y_c$. Notice that the more general solution has the same behaviour at $x\to\pm\infty$ and, hence, it has the same values of the conserved quantities. In particular, since the two constituents can be well separated, this is consistent with the mass being additive 
\EQ{
M_{ab}= M_{ac}+ M_{cb}\ .
}
So, if we order the eigenvalues, say $m_1<m_2<\cdots$, then the elementary soliton solutions are only associated to the pairs $\Bvarpi=\Bv_a+\Bv_{a+1}$.
Below we consider some examples:

\noindent
{\bf(AIII)} There are three distinct classes of soliton solutions:

(a) The first has dressing data
\EQ{
 \{\xi,-\xi\}\ ,\qquad\{\Bv_a^++\BOmega,\Bv_a^-+\BOmega\}\ ,
\label{pqq1}
 }
 where we used \eqref{swop}, along with $I_{np}\BOmega=\BOmega$. According to the discussion above, only the soliton with $a=1$ will be elementary.
From \eqref{topc}, the kink charge is
\EQ{
q_0=-2iq\>  \left(2\BOmega\BOmega^\dagger- \Bv^+_a{\Bv_a^+}^\dagger-\Bv^-_a{\Bv_a^-}^\dagger\right)\ .
\label{topc12}
}
Notice that $q_0$ is actually ambiguous up to shifts $q\to q+\pi$ since the kink charge only appears via $e^{q_0}$. In fact, for future use it will be convenient to define
\EQ{
\bar q=\begin{cases}q & 0<q\leq\frac\pi2\\ q-\pi & \frac\pi2\leq q<\pi\ ,\end{cases}
}
so that $-\frac\pi2\leq\bar q\leq\frac\pi2$ and the topological charge is equivalently
\EQ{
q_0=-2i\bar q\>  \left(2\BOmega\BOmega^\dagger- \Bv^+_a{\Bv_a^+}^\dagger-\Bv^-_a{\Bv_a^-}^\dagger\right)\ .
\label{topc2}
}
It is then natural to think of solutions with $\bar q>0$ as solitons and $\bar q<0$ as anti-solitons with charge conjugation taking $\bar q\to-\bar q$. 
The mass is
\EQ{
M=\frac{4k}\pi\left|m_a\sin \bar q\right|\ ,
\label{mss}
}
The solution has internal collective coordinates corresponding to the choice of the unit vector $\BOmega$ in the degenerate null subspace. The moduli space has the form of a co-adjoint orbit of $SU(n-p)\subset H$ given by 
\EQ{
q_0= -2i\bar q\> U \left( 2E_{11} - E_{n-p+a, n-p+a} -E_{n+a,n+a}\right) U^{-1}\ , 
}
where
\EQ{
U\equiv 
\left(\begin{array}{c|c|c}
U_{n-p}&&\\ \hline
&{\mathbb I}_{p}&\\ \hline
&&{\mathbb I}_{p}
\end{array}\right) \in SU(n-p)\subset SU(n+p)\ .
}
This is the action of $U\in H$ on the soliton.
More schematically, the orbit is of the form 
\EQ{
q_0=-4i\bar q\,U\MAT{1&&\\ &0&\\&&\ddots}U^{-1}\ ,\qquad U\in SU(n-p)
\label{orb1}
}
and so is identified with the co-adjoint orbit
\EQ{
{\mathfrak M}=\frac{SU(n-p)}{U(n-p-1)}\simeq\CP^{n-p-1}\ . 
}

(b) The second kind of soliton has dressing data
\EQ{
 \{\xi,-\xi\}\ ,\qquad\{\Bv_a^++\Bv_{b}^+,\Bv_a^-+\Bv_{b}^-\}\ , 
 }
 with $a\neq b=1,\ldots,p$, although only the cases $b=a\pm1$ will be elementary.
The kink charge is 
\EQ{
q_0=-2i\bar q\left(\Bv^+_{a}{\Bv_{a}^+}^\dagger+\Bv^-_{a}{\Bv_{a}^-}^\dagger- \Bv^+_{b}{\Bv_b^+}^\dagger-\Bv^-_{b}{\Bv_{b}^-}^\dagger\right)\ .
\label{topc3}
}
and the mass is
\EQ{
M=\frac{4k}{\pi}\left|(m_a-m_b)\sin\bar q\right|\ .
\label{mss2}
}
In this case the kink charge is associated to the abelian subgroup $U(1)^p\subset H$ and the internal moduli space is trivial.

(c) The third kind of soliton has dressing data
\EQ{
 \{\xi,-\xi\}\ ,\qquad\{\Bv_a^++\Bv_{b}^-,\Bv_a^-+\Bv_{b}^+\}\ , 
 }
 with $a,b=1,\ldots,p$, although none of these will be elementary. 
The kink charge is 
\EQ{
q_0=-2i\bar q\left(\Bv^+_{a}{\Bv_{a}^+}^\dagger+\Bv^-_{a}{\Bv_{a}^-}^\dagger- \Bv^+_{b}{\Bv_b^+}^\dagger-\Bv^-_{b}{\Bv_{b}^-}^\dagger\right)\ .
\label{topc41}
}
and the mass is
\EQ{
M=\frac{4k}{\pi}\left|(m_a+m_b)\sin\bar q\right|\ .
\label{mst3}
}

\noindent
{\bf(BDI)}  As for the complex case, there are 3 kinds of solitons:

(a) The first one is specified by the dressing data
\EQ{
\{\xi,\xi^*,-\xi,-\xi^*\}\ ,\qquad
\big\{\Bv_a^++\BOmega,\Bv_a^-+\BOmega^*,
\Bv_a^-+\BOmega,\Bv_a^++
\BOmega^*\big\}\ .
}
The additional constraint \eqref{extc} requires $\BOmega\cdot\BOmega=0$, and it is useful to 
write
\EQ{
\BOmega=\frac1{\sqrt{2}}\big(\BOmega^{(1)}-i\BOmega^{(2)}\big)\ ,
}
where $\BOmega^{(1,2)}$ are two real unit vectors with
\EQ{
\BOmega^{(1)}\cdot\BOmega^{(2)}=0\ .
}
Hence, the kink charge is
\EQ{
q_0=-4i\bar q\big(\BOmega\BOmega^\dagger-\BOmega^*\BOmega^t\big)=4\bar q(\BOmega^{(1)}{\BOmega^{(2)}}^t-\BOmega^{(2)}{\BOmega^{(1)}}^t)\ ,
\label{orb2}
}
where we used the fact that $I_{np}\Bv_a^+=\Bv_a^-$ and ${\Bv^+_a}^*=\Bv_a^-$.
The mass of the soliton is
\EQ{
M=\frac{4k}\pi\left|m_a\sin \bar q\right|\ .
}
As for the complex Grassmannians, only the soliton with $a=1$ will be elementary.

This soliton has an internal moduli space which is a co-adjoint 
orbit of $SO(n-p)$ given by 
\EQ{
q_0= -4\bar q\> U \left( E_{21} - E_{12} \right) U^{-1}\ .
}
Schematically this is of the form
\EQ{
q_0=4\bar q\,U\MAT{0&-1&&&\\ 1&0&&\\ &&0&\\ &&&\ddots}U^{-1}\ ,\qquad U\in SO(n-p)\ ,
}
which has the form of a real Grassmannian 
\EQ{
{\mathfrak M}=\frac{SO(n-p)}{SO(2)\times SO(n-p-2)}\ .
}
Since the vector representation of $SO(n-p)$ is real, there exists an element of $SO(n-p)$ which reverses the sign of $\bar q$. Therefore, in this case, we can further restrict $\bar q$ to the range $0< \bar q\leq\frac\pi2$, because solutions with $\bar q<0$ lie in the co-adjoint orbit of a soliton with $\bar q'=|\bar q|$.

(b) The second kind of soliton has dressing data
\EQ{
 \{\xi,\xi^*,-\xi,-\xi^*\}\ ,\qquad\{\Bv_a^++\Bv_{b}^+,\Bv_a^-+\Bv_{b}^-,\Bv_a^-+\Bv_{b}^-,\Bv_a^+ +\Bv_{b}^+\}\ ,
 }
for $a\not=b=1,\ldots, p-1$, although only the ones with $b=a\pm1$ are elementary. In these cases the kink charge vanishes and for $b=a+1$ the mass is
\EQ{
M=\frac{4k}\pi\left|(m_{a+1}-m_a)\sin \bar q\right|\ .
\label{mss5}
}
The moduli space is trivial, and the soliton carries no $H$-charges.

(c) The third kind of soliton has dressing data
\EQ{
 \{\xi,\xi^*,-\xi,-\xi^*\}\ ,\qquad\{\Bv_a^++\Bv_{b}^-,\Bv_a^-+\Bv_{b}^+,\Bv_a^-+\Bv_{b}^+\Bv_a^-+\Bv_{b}^+\}\ , 
 }
with $a\not=b=1,\ldots,p$, although none of these will be elementary. 
The kink charge vanishes,
and the mass is
\EQ{
M=\frac{4k}{\pi}\left|(m_a+m_b)\sin\bar q\right|\ .
\label{mst2}
}

\noindent
{\bf (CII)} Once again there are three classes of soliton solution:

(a) The solitons with non-abelian $H$ charges are associated to the dressing data 
\EQ{
\{\xi,\xi^*,-\xi,-\xi^*\}\ ,\qquad
\big\{\Bv_a^++\BOmega,J_{n+p}(\Bv_a^-+\BOmega^*),
\Bv_a^-+\BOmega,J_{n+p}(\Bv_a^++
\BOmega^*)\big\}\ ,
}
with kink charge
\EQ{
q_0&=4i\bar q\big(\BOmega\BOmega^\dagger-J_{n+p}\BOmega^*\BOmega^tJ_{n+p}^{-1}
- \Bv^+_a{\Bv_a^+}^\dagger-\Bv^-_a{\Bv_a^-}^\dagger\\
&~~~~~~\qquad+J_{n+p}\Bv^+_a{\Bv_a^+}^\dagger J_{n+p}^{-1}+J_{n+p}\Bv^-_a{\Bv_a^-}^\dagger J_{n+p}^{-1}
\big)
}
and mass is
\EQ{
M=\frac{4k}\pi\left|m_a\sin \bar q\right|\ .
\label{mss4}
}
Once again only the soliton with $a=1$ is elementary.

This soliton has an internal moduli space which is a co-adjoint 
orbit of the subgroup $Sp(n-p)\times Sp(1)\subset H$ of the form
\EQ{
{\mathfrak M}=\frac{Sp(n-p)}{Sp(n-p-1)\times U(1)}\times\frac{Sp(1)}{U(1)}\ .
}
In this case, there exists an element of $H$ which can reverse the sign of 
$\bar q$ and so we can restrict  $0<\bar q\leq\frac\pi2$. 

(b) The second kind of soliton has dressing data 
\EQ{
 \{\xi,\xi^*,-\xi,-\xi^*\}\ ,\qquad\{\Bv_a^++\Bv_{b}^+,J_{n+p}(\Bv_a^-+\Bv_{b}^-),\Bv_a^-+\Bv_{b}^-,J_{n+p}(\Bv_a^+ +\Bv_{b}^+)\}\ ,
 }
for $a=1,\ldots, p-1$. The mass is
\EQ{
M=\frac{4k}\pi\left|(m_{b}-m_a)\sin \bar q\right|\ .
\label{mss3}
}
Only the solitons with $b=a\pm1$ is elementary and the moduli space is a product
\EQ{
{\mathfrak M}=\frac{Sp(1)}{U(1)}\times\frac{Sp(1)}{U(1)}\ .
}

(c) The third kind of soliton has dressing data
\EQ{
 \{\xi,\xi^*,-\xi,-\xi^*\}\ ,\qquad\{\Bv_a^++\Bv_{b}^-,J_{n+p}(\Bv_a^-+\Bv_{b}^+),\Bv_a^-+\Bv_{b}^+,J_{n+p}(\Bv_a^+ +\Bv_{b}^-)\}\ ,
 }
for $a=1,\ldots, p-1$. The mass is
\EQ{
M=\frac{4k}\pi\left|(m_{b}+m_a)\sin \bar q\right|
\label{mss33}
}
None of these solutions is elementary.

\noindent
{\bf(AII)} In this case, solitons are associated to the dressing data
\EQ{
\{\xi,-\xi^*\}\ ,\qquad\{\BOmega_a+\BOmega_b,J_n(\BOmega_a^*+\BOmega_b^*)\}\ , \qquad a\not=b\ ,
}
where $\BOmega_a$ and $\BOmega_b$ are the generic eigenvectors defined in~\eqref{EigenVAII}.
The solitons have a mass $M=(2k/\pi)|(m_b-m_a)\sin\bar q|$ and kink charge
\EQ{
q_0&=2i\bar q\left(\BOmega_{a}\BOmega_{a}^\dagger-J_n\BOmega_{a}^*\BOmega_{a}^tJ_n^{-1}- 
\BOmega_{b}\BOmega_{b}^\dagger+J_n\BOmega_{b}^*\BOmega_{b}^tJ_n^{-1}\right)\\ &
=2i\bar q\> U\big(\Be_a\Be_a^t-\Be_{a+n}\Be_{a+n}^t-\Be_{b}\Be_{b}^t+\Be_{b+n}\Be_{b+n}^t\big)U^{-1}
\ .
\label{topc4}
}
Here, $U=U_aU_b$, and $U_a$ and $U_b$ are particular elements of two mutually commuting $SU(2)$ subgroups or $H$ fixed by the conditions $\BOmega_a = U_a \Be_a$ and $\BOmega_b = U_b \Be_b$. 
In other words, these solitons have an internal moduli space which is a co-adjoint orbit of the product group $SU(2)\times SU(2)\subset H$, schematically of the form
\EQ{
q_0=2i\bar q\,U\MAT{1&0\\ 0&-1}U^{-1}
\label{efa}
}
for each $SU(2)$ factor. This identifies it as
\EQ{
{\mathfrak M}=\frac{SU(2)}{U(1)}\times\frac{SU(2)}{U(1)}\ .
}
Only the solitons with $b=a\pm 1$ will be elementary.

\noindent
{\bf Non-generic $\Lambda$}

We will not consider the case of non-generic $\Lambda$ in much detail, since the generalization is for the most part obvious; rather we will take the case of the complex Grassmannians AIII as an illustrative example.
In general, the spectrum of $\Lambda$ will be of the form~\footnote{For the case where $\Lambda$ is regular, $d_i=1$ for each $i=1,\ldots,p$.}
\EQ{
\big\{-im_p^{d_p},\ldots,-im_1^{d_1},0^{n+p-2\sum_id_i},im_1^{d_1},\ldots,im_p^{d_p}\big\}\ ,
}
where some of the integers $d_1,\ldots, d_p$ may vanish. In this case
\EQ{
H=S\big(U(d_0)\times U(d_1)\times\cdots U(d_p)\big)\supset S(U(n-p)\times U(1)^p)=H_\text{regular}\ ,
}
with $d_0=n+p-2\sum_{i=1}^pd_i\geq n-p$.
If we maintain the ordering $m_a<m_{a+1}$, then elementary solitons will be associated to pairs $(a,a+1)$ with masses as in \eqref{mss2} (allowing $a=0$ with $m_0=0$). This soliton has an internal moduli space which is the product of two co-adjoint orbits:
\EQ{
{\mathfrak M}=\frac{SU(d_a)}{U(d_a-1)}\times\frac{SU(d_{a+1})}{U(d_{a+1}-1)}\ .
}

\section{Semi-Classical Quantization}
\label{SMquantum}

In this section, we describe how to quantize the internal
degrees-of-freedom of the elementary solitons in the semi-classical approximation.
The internal moduli space is identified with the orbit
\EQ{
\Bvarpi\longrightarrow U\Bvarpi\ ,
\label{collc}
}
for $U$ a constant element of $\tilde H$, a semi-simple subgroup of $H$. 
The classical moduli space is
identified with the quotient ${\mathfrak M}=\tilde H/H_0$, where $H_0\subset\tilde H$ is the stability group of the solution. We can also think of the moduli space as an adjoint orbit,
\EQ{
\Bvarpi\Bvarpi^\dagger\longrightarrow U\Bvarpi\Bvarpi^\dagger U^\dagger\ .
}
It is well-known that this
(co-)adjoint orbit is a homogeneous symplectic manifold and can be
quantized. The term homogeneous refers to the fact that the symplectic form is invariant under $\tilde H$ which acts 
transitively  on ${\mathfrak M}$ by left group action. In fact, due to a theorem of Borel \cite{Borel}, the co-adjoint orbits of a compact group are actually homogeneous K\"ahler manifolds.\footnote{A nice physicist's review of these spaces appears in \cite{Bordemann:1985xy}.}
However, we have to show how this structure actually arises in the SSSG theory. Note that the cases that arise for the solitons are the orbits of the form
\EQ{
\frac{SU(m)}{U(m-1)}\ ,\qquad\frac{SO(m)}{SO(m-2)\times SO(2)}\ ,\qquad
\frac{Sp(m)}{Sp(m-1)\times U(1)}\ ,
}
or a product of at most two of these of the same kind. 

The techniques for semi-classically quantizing a soliton with an internal moduli space are well known in many other situations. The main idea, originating with Manton \cite{Manton:1981mp}, is to construct an effective mechanical theory of the moduli space dynamics along the world-line of the soliton which can then be quantized. In this process, we focus only on the internal collective coordinates and ignore the collective coordinates corresponding to the position of the soliton, which are trivially quantized. 
To this end, we allow the collective coordinates to 
become time-dependent by taking $U\to U(t)$ in \eqref{collc}. Of course, the soliton solution with time-dependent collective coordinates is not a solution of the equations-of-motion. Hence, the solution has to be modified to take account of the back-reaction.
When the back-reacted solution is substituted into the action of the theory one will obtain a non-trivial functional of $U(t)$ which can be expanded in powers of derivatives $dU/dt$. What results is a finite dimensional Lagrangian system on the orbit ${\mathfrak M}=\tilde H/H_0$.
Fortunately, to the leading-order semi-classical approximation, one only needs the terms with the minimal number of derivatives, and this is obtained by taking the raw, un-back-reacted, soliton solution with $U\to U(t)$, substituting it into the action and taking the term with the minimum number of derivatives. In many familiar situations, for example for monopoles in gauge theory in $3+1$ dimensions, these terms are quadratic in time derivatives. However, in the present context, we will find that there are terms linear in derivatives and keeping these will give the leading-order semi-classical approximation. 

In order to find the effective finite dimensional Lagrangian, we substitute
\EQ{
\gamma\longrightarrow U(t)\gamma\, U(t)^{-1}
}
into the action of the theory. It is important to realize that this is not a gauge transformation because the gauge field, which vanishes on-shell for a soliton solution, is fixed at $A_\mu=0$.
In Appendix~A, we prove that the 
effective quantum mechanical action for the collective coordinates of the soliton to lowest order in the expansion in $t$ derivatives involves simply the kink charge of the soliton (eq.~\eqref{TheMissingLink})
\EQ{
S_\text{eff}[U]=\int_{-\infty}^\infty dt\,\Tr\left(U^{-1}\frac{dU}{dt}\,\sigma\right)\ ,\qquad \sigma=\frac{\kappa}{2\pi}q_0\ .
\label{cvc2}
}
The quantum mechanical system defined by this Lagrangian is in the class of Chern-Simons Quantum Mechanics defined in \cite{Ivanov:2003qq} since we can think of the action as $S=\int A$, where $A$ is the pull-back of a 1-form on the target space ${\mathfrak M}$ to the world-line of the soliton. These kinds of quantum mechanical systems are topological in the sense that they do not depend on the metric of ${\mathfrak M}$. We will see that they are quite simple and are solved by imposing a consistent set of constraints. It is also worth remarking that the internal 
dynamics does not contribute to the mass of the soliton since the Hamiltonian for the internal motion vanishes. In the following, we develop two approaches to the problem, the first is more direct and motivated by the parameterization of the co-adjoint orbit provided by the soliton itself. In this approach, the non-commutative or fuzzy geometric version of $\mathfrak M$ emerges directly. The second approach is 
Lie algebraic and necessarily rather abstract but more general. Of course both approaches are completely equivalent.

\subsection{The direct approach}

(a) ${\mathfrak M}=SU(m)/U(m-1)\simeq\CP^{m-1}$, which arises in the case of the complex Grassmannians (and also other cases for $m=2$). The internal collective coordinate of the soliton is the normalized complex $m$ vector $\BOmega$. In fact, this provides a parameterization of the orbit in the form \eqref{orb1}\footnote{This follows by taking $\BOmega(t)=U(t)\BOmega_0$, where $\BOmega_0$ is the fixed reference point with $\sigma=(2i\bar qk/\pi)\BOmega_0\BOmega_0^\dagger$.}
\EQ{
U(t)\sigma U(t)^{-1}=\frac{2i\bar qk}\pi\,\BOmega(t)\BOmega(t)^\dagger\ .
}
The orbit ${\mathfrak M}$ is defined by the unit vector $\BOmega$ modulo multiplication by a phase. Notice that $\BOmega$ transforms as a vector under  $SU(m)$, which we identify with the left action on the homogeneous space ${\mathfrak M}$. In terms of this variable, the action \eqref{cvc2} takes the simple form ($\kappa=k$)
\EQ{
S_\text{eff}=-\frac{2i\bar qk}{\pi}\int dt\,\BOmega^*\cdot\frac{d\BOmega}{dt}\ .
}

It is useful to 
enlarge the phase space to ${\mathbb C}^m$ since then the Poisson brackets are simpler: 
\EQ{
\{\BOmega_i,\BOmega^*_j\}=\frac{i\pi}{2\bar qk}\delta_{ij}\ .
} 
The way to reduce the larger phase space to $\CP^{m-1}$ involves a K\"ahler quotient. This starts by noticing that 
the $U(1)$ symmetry $\BOmega\to e^{i\alpha}\BOmega$ 
is a Hamiltonian symmetry generated by 
$\Phi=\BOmega^*\cdot\BOmega$. The physical phase space corresponds to restricting ${\mathbb C}^m$ to the level set 
\EQ{
\Phi=\BOmega^*\cdot\BOmega=1
}
and performing a quotient by the $U(1)$ symmetry which yields $\CP^{m-1}$ directly.

In the quantum theory, we can replace the Poisson brackets by commutators 
involving the operators $\hat\BOmega_i$ and $\hat\BOmega_i^\dagger$:
\EQ{
[\hat\BOmega_i,\hat\BOmega_j^\dagger]=\frac{\pi}{2\bar qk}\delta_{ij}
\label{nco}
}
and build a Hilbert space by treating the former as annihilation operators and the latter as creation operators for solitons $\bar q>0$. For the anti-solitons $\bar q<0$ the r\^oles of the operators are interchanged.
The generator of the Hamiltonian symmetry 
\EQ{
\hat\Phi=\hat\BOmega^\dagger\cdot\hat\BOmega =\frac{\pi}{2\bar qk}\hat{\EuScript N}
}
is proportional to the number operator (which we take to be negative for the anti-solitons), and the constraint $\hat\Phi=1$, along with the quantization of the occupation number, implies the quantization of $\bar q$:
\EQ{
\bar q=\frac{\pi {\EuScript N}}{2k}\ ,\qquad {\EuScript N}=\pm1,\pm2,\ldots,\pm k\ ,
\label{iuu}
}
where we have taken account of the fact that  $-\frac\pi2\leq\bar q\leq\tfrac\pi2$. For solitons, the Hilbert space is spanned by the states\footnote{Notice that the quotient by $U(1)$ is trivial at the level of the Hilbert space.}
\EQ{
\hat\BOmega_{i_1}^\dagger\hat\BOmega_{i_2}^\dagger\cdots\hat\BOmega_{i_{\EuScript N}}^\dagger|0\rangle\ ,
}
which identifies it as the representation space for the rank-${\EuScript N}$ symmetric representation of $SU(m)$. The anti-solitons transform in the conjugate rank-$\EuScript N$ symmetric representations.

The construction we have presented has an interesting interpretation as ``fuzzy geometry''. This follows from the fact that  in the quantum theory the coordinates of $\CP^{n-1}$ do not commute as in \eqref{nco}.
However, as $|{\EuScript N}|$ increases the non-commutativity 
gets less marked and the fuzzy geometry 
becomes a closer approximation of the classical geometry in the limit $|{\EuScript N}|\to\infty$, which clearly requires $k\to\infty$, the semi-classical limit. In the next section, we identify the quasi-classical states in this limit.

(b) ${\mathfrak M}=SO(m)/SO(2)\times SO(m-2)$, which arises in the case of the real  Grassmannians. 
The internal collective coordinate of the soliton is the unit complex $m$ vector $\BOmega$ which is subjected to the constraint $\BOmega\cdot\BOmega=0$. In fact, this provides a parameterization of the orbit in the form \eqref{orb2} ($\kappa=k/2$)
\EQ{
U(t)\sigma U(t)^{-1}=\frac{i\bar qk}\pi\,\left(\BOmega(t)\BOmega(t)^\dagger-\BOmega(t)^*\BOmega(t)^t\right)\ .
}
In this case, the effective action takes the form 
\EQ{
S_\text{eff}=-\frac{i\bar qk}{\pi}\int dt\,\Big(\BOmega^*\cdot\frac{d\BOmega}{dt}-\BOmega\cdot\frac{d\BOmega^*}{dt}\Big)=-\frac{2iqk}{\pi}\int dt\,\BOmega^*\cdot\frac{d\BOmega}{dt}\ .
}
However, there is the additional constraint $\BOmega\cdot\BOmega=0$ to impose. Since this is holomorphic
it can be implemented directly at the level of the Fock space, which is the one of case (a), by removing by hand states which are of the form $\sum_i\cdots\hat\BOmega_i^\dagger\cdots\hat\BOmega_i^\dagger\cdots |0\rangle$.  One recognizes this as the process of ``removing traces" that is well-known in the Young Tableaux approach to the orthogonal groups. The remaining states form a representation space for the symmetric representation of $SO(m)$. The quantization condition now becomes
\EQ{
\bar q=\frac{\pi {\EuScript N}}{2k}\ ,\qquad {\EuScript N}=1,2,\ldots,k\ .
\label{iuu2}
}
In this case, the solitons and anti-solitons lie in the same representations, since the representations are real, and $\bar q$ is restricted to $0<\bar q\leq\frac\pi2$. 

(c) ${\mathfrak M}=Sp(m)/Sp(m-1)\times U(1)$, which arises in the case of the quaternionic Grassmannians. 
The internal collective coordinate of the soliton is the unit complex $2m$ vector $\BOmega$. In fact, this provides a parameterization of the orbit in the form  ($\kappa=k/2$)
\EQ{
U(t)\sigma U(t)^{-1}=\frac{i\bar qk}\pi\,\left(\BOmega(t)\BOmega(t)^\dagger-J_m\BOmega(t)^*\BOmega(t)^tJ_m^{-1}\right)\ .
}
In this case the the effective action takes the form 
\EQ{
S_\text{eff}=-\frac{i\bar qk}{\pi}\int dt\,\Big(\BOmega^*\cdot\frac{d\BOmega}{dt}-\BOmega\cdot\frac{d\BOmega^*}{dt}\Big)=-\frac{2i\bar qk}{\pi}\int dt\,\BOmega^*\cdot\frac{d\BOmega}{dt}\ ,
}
the equivalence being up to total derivatives. The Fock space is constructed as in the case (a) with $m\to2m$ and the states transform in the symmetric representations of $Sp(m)$. The quantization condition is
\EQ{
\bar q=\frac{\pi {\EuScript N}}{2k}\ ,\qquad {\EuScript N}=\pm1,\pm2,\ldots,\pm k\ .
\label{iuu3}
}
In this case solitons and anti-solitons transform in the same real representations; however, as mentioned earlier they have opposite abelian charges.

Notice that the quantization condition on $\bar q$ takes a universal form for all three cases.

\subsection{Lie algebraic approach}

We will follow the approach taken in \cite{Alexanian:2001qj} (see also \cite{Ivanov:2003qq}) for
quantizing the co-adjoint orbit for the example 
${\mathfrak M}=SU(3)/U(2)$ and extend it to an arbitrary co-adjoint orbit. This approach has the advantage of being able to easily extend to arbitrary groups. First of all, we introduce 
local coordinates on $U\in\tilde H$ by writing
$U=e^{\lambda_i\theta^i}$, where $\{\lambda_i\}$ 
is a basis of normalized anti-hermitian generators of the Lie algebra $\tilde{\mathfrak h}$,
\EQ{
\Tr\,(\lambda_i\lambda_j)=-\delta_{ij}\ .
}
It is then useful to define $E_{ij}(\theta)$ via the left invariant (Maureen-Cartan) 1-forms
$U^{-1}dU=\lambda_iE_{ij}d\theta^j$. Note that because we normalized the generators as above there is no real distinction between upper and lower algebra indices.

The co-adjoint orbit is a symplectic manifold where the symplectic 2-form, the Kirillov-Konstant 2-form, is
\EQ{
\Omega=dA=-\Tr\left(\left(U^{-1}dU\wedge U^{-1}dU\right)\sigma\right)\ . 
\label{sym}
}
Notice that the symplectic form is invariant under left multiplication by the group $U\to VU$, $V\in\tilde H$.
In order to quantize the system, it is useful to first enlarge the phase space
by introducing Lagrange multipliers and constraints
such that
\EQ{
S_\text{eff}&=\int dt\,\Tr\left(U^{-1}\frac{dU}{dt}\,\sigma\right)= \int dt\,E_{ji} \Tr(\sigma \lambda_j) \frac{d\theta^i}{dt}\\ &~~~  \;\longrightarrow\; S_\text{eff}=\int dt\,\Big(p_i\frac{d\theta^i}{dt}+\ell^i\varphi_i\Big)\ ,
\label{cvc3}
}
where the $\ell^i$ are the Lagrange multipliers and the constraints are
\EQ{
\varphi_i=p_i-E_{ji}\Tr(\sigma\lambda_j)\approx 0\ .
\label{cons}
}
The enlarged phase space has simple Poisson brackets $\{\theta^i,p_j\}=\delta^i{}_j$
and the non-trivial nature of the geometry is now encoded in the constraints. 
It is useful to define the quantities,
\EQ{
\Lambda_a=p_j(E^{-1})_{ji}\Tr(a\lambda_i)\ ,
}
for $a\in{\mathfrak h}$. These 
generate the right action of the group $\tilde H$ and have Poisson brackets 
\EQ{
\{\Lambda_a,U\}=Ua\
,\qquad\{\Lambda_a,\Lambda_b\}=\Lambda_{[a,b]}\ .
}
In terms of these variables the constraints
\eqref{cons} become
\EQ{
\Lambda_a\approx\Tr\left(a\sigma\right)\ .
\label{cons2}
}
Now we decompose the Lie algebra $\tilde{\mathfrak h}$ into that of the
Lie algebra of the stability group $H_0\subset\tilde H$, that is the subgroup for
which $U\sigma U^{-1}=\sigma$, and its complement
\EQ{
\tilde{\mathfrak h}={\mathfrak h}_0\oplus{\mathfrak r}\ .
}
Furthermore, we can separate out of ${\mathfrak h}_0$ the abelian factor generated by $\sigma$:
\EQ{
{\mathfrak
  h}_0={\mathfrak h}_\sigma\oplus{\mathfrak h}'_0\ .
}
Then because the Lie algebra has a schematic structure
\EQ{
[{\mathfrak h}_0,{\mathfrak h}_0]=
{\mathfrak h}'_0\ ,\qquad [{\mathfrak h}_0,{\mathfrak r}]={\mathfrak
   r}\ ,\qquad[{\mathfrak r},{\mathfrak r}]={\mathfrak r}\oplus{\mathfrak h}_\sigma\oplus 
{\mathfrak h}'_0\ ,
}
the constraints \eqref{cons2}, for $\Lambda_a$ with $a\in{\mathfrak h}_0$, are, in the
language of constrained Hamiltonian systems, first
class constraints, while those for $\Lambda_a$ with $a\in{\mathfrak r}$, are second
class. The first class constraints can be imposed directly on the
phase space.
In order to be concrete, let us introduce the usual basis of
generators for the complexified Lie
algebra ${\mathfrak h}$, $\{\vec H,E_{\vec\alpha}\}$, where
$\vec\alpha$ lies in the set of roots $\Phi$.
Without loss of generality, we can assume that $\sigma$ lies in the Cartan subalgebra $\sigma=i\vec\sigma\cdot
\vec H$ for a vector
$\vec\sigma$.\footnote{The Cartan generators $\vec H$ are realized as a rank$({\mathfrak h})$ vector with $\vec H^\dagger=\vec H$.} Furthermore, we shall fix the positive roots of
${\mathfrak h}$ relative to $\vec\sigma$; in other words, for a
positive root $\vec\alpha\cdot\vec\sigma\geq0$.
The Lie algebra of the stability group ${\mathfrak h}_0$
is then identified with $\{\vec H,E_{\vec\alpha}\}$, with
$\vec\alpha\cdot\vec\sigma=0$. 
This means that
the first class constraints take the form
\EQ{
\Lambda_{\vec H}\approx\vec\sigma\ ,\qquad \Lambda_{E_{\vec\alpha}}\approx0\qquad\text{if}\qquad\vec\alpha\cdot\vec\sigma=0\ .
}

The second class constraints can be dealt with by using the machinery of Dirac brackets. However, here we follow both \cite{Ivanov:2003qq} and \cite{Alexanian:2001qj}, and exploit the fact that ${\mathfrak M}$ is actually a K\"ahler manifold. The idea to consider is to take appropriate complex combinations of the constraints and separate them into two mutually complex conjugate sets. The constraints within each set mutually Poisson commute, and so one of the sets can be chosen to be effectively first class constraints which can be treated in the usual way. When implemented in the quantum theory this procedure is known as ``analytic quantization" \cite{Ivanov:2003qq}.
The appropriate complex combinations correspond to taking $\Lambda_{E_{\pm\vec\alpha}}$ for
$\vec\alpha$ a root for generators in ${\mathfrak r}$.
If $\bar q>0$ ($\bar q<0$) the appropriate choice of constraints are
$\Lambda_{E_{\vec\alpha}}$ for $\vec\alpha$ a positive (negative)
root, which we write as $\vec\alpha>0$ ($\vec\alpha<0$). It is
convenient to write the set of roots of $\tilde{\mathfrak h}$ as
$\Phi=\Phi_0\cup\Phi^{\mathfrak r}_+\cup\Phi^{\mathfrak r}_-$, where $\Phi_0$ are the 
roots of
${\mathfrak h}_0$ and $\Phi^{\mathfrak r}_\pm$ are the positive, respectively,
negative, roots of ${\mathfrak r}$. 
The resulting first class constraints are then 
\EQ{
\Lambda_{E_{\vec\alpha}}\approx 0\qquad\text{for}\qquad\vec\alpha\in\Phi^{{\mathfrak r}}_{\text{sign}(\bar q)}\ .
}
To summarize, the final set of what are effectively first class constraints is
\EQ{
\Lambda_{\vec H}\approx\vec\sigma\ ,\qquad
\Lambda_{E_{\vec\alpha}}\approx 0\qquad\text{for}\qquad\vec\alpha\in\Phi_0\cup
\Phi_{\text{sign}(\bar q)}^{\mathfrak r}\ .
\label{gfg}
}

Now we quantize the system by replacing Poisson brackets by
commutators $\{\cdot,\cdot\}\to i[\cdot,\cdot]$ with
wavefunctions on the group $\psi(U)$, $U\in\tilde H$. 
The Peter-Weyl Theorem guarantees that a basis for the Hilbert space of the enlarged system
is obtained by taking wavefunctions which are just the matrix elements of
$U\in\tilde H$ in all possible irreducible representations of the group,
that is of the form
\EQ{
\psi(U)=\langle\rho_1|U|\rho_2\rangle\ ,
}
where $U$ is taken in an irreducible representation and $|\rho_i\rangle$ are
two elements of the associated module. The classical variables
$\Lambda_a$ are then lifted to quantum operators $\hat\Lambda_a$ which
implement right multiplication by the corresponding element of the Lie
algebra: 
\EQ{
\hat\Lambda_a\psi(U)=\langle\rho_1|Ua|\rho_2\rangle\ .
}
Imposing the constraints
\eqref{gfg} on the Hilbert space implies that in the representation
$R$ we must have
\EQ{
\vec H|\rho_2\rangle=\vec\sigma|\rho_2\rangle\ ,\qquad
E_{\vec\alpha}|\rho_2\rangle=0~~~~
\vec\alpha\in\Phi_0\cup\Phi_{\text{sign}(\bar q)}^{\mathfrak r}\ .
\label{aza}
}
In other words, $|\rho_2\rangle$ 
must be a highest (lowest) weight state
for $q>0$ ($q<0$). The requirement \eqref{aza} implies the quantization condition
on $q$ such that
\EQ{
\vec\sigma\in\text{weight lattice}\ .
}
The kink charge takes the form $q_0=i\bar q\vec\gamma\cdot\BH$, then  
if $\ell$ is the smallest positive real number such that $\ell\vec\gamma$ is a weight vector, then $q$ is quantized as
\EQ{
\bar q=\frac{2\pi\ell}{\kappa}\, {\EuScript N}\ ,\qquad{\EuScript N}=\pm1,\pm2,\cdots,\pm\left[\frac{\kappa}{4\ell}\right]\ .
\label{qcon}
}
The states are then restricted to the irreducible representations
of $\tilde H$ with a highest (lowest) weight $\vec\lambda={\EuScript N}
\ell\vec\gamma$, for ${\EuScript N}\in{\mathbb Z}>0$
(${\EuScript N}\in{\mathbb Z}<0$) of the form
\EQ{
\langle \rho|U|\vec\lambda\rangle\ ,
}
with $|\vec\lambda\rangle$ 
the highest (lowest) weight state of the representation and $|\rho\rangle$
an arbitrary state in the representation space. Consequently, we can identify the
Hilbert space with the representation of $\tilde H$ with highest (lowest)
weight $\vec\lambda$, for ${\EuScript N}>0$ (${\EuScript N}<0$). These modules are representation spaces for the left action of $\tilde H$.

In the representation space one can construct a set of coherent states of the form \cite{Perelomov,Perelomov:1971bd}
\EQ{
|p(U);{\EuScript N}\rangle\equiv|\rho\rangle=U|\vec\lambda\rangle\ .
}
The inequivalent states of the this form are labelled by  points in the orbit $p(U)\in{\mathfrak M}=\tilde H/H_0$ since group elements $U\in H_0$ leave the highest (lowest) weight vector invariant. For finite ${\EuScript N}$ these states are, obviously, overcomplete; however, in the classical limit as ${\EuScript N}\to\infty$ they can be thought of as quasi-classical states that approximate the classical configuration $p(U)\in {\mathfrak M}$. On the contrary, when ${\EuScript N}$ is finite we have a ``quantum" or ``fuzzy" geometry.

Below we consider the 3 cases of interest to us:

(a) ${\mathfrak M}=SU(m)/U(m-1)\simeq\CP^{m-1}$. The orbit takes the form
\EQ{
U\vec\gamma\cdot\vec H U^{-1}=4\BOmega\BOmega^\dagger\ ,
}
which means that
 \EQ{
\vec\gamma=4\vec\omega_1\ ,
}
where $\vec\omega_1$ is the highest weight of the defining $m$-dimensional vector representation of $SU(m)$. Consequently, in this case $\ell=\frac14$ and the quantization condition \eqref{qcon} is precisely as in \eqref{iuu}.
The Hilbert space consists of the representations
with highest weight ${\EuScript N}\vec\omega_1$, for ${\EuScript N}=1,2,\ldots,k$, 
that is the symmetric representations, and those with lowest weight
${\EuScript N}\vec\omega_1$, for ${\EuScript N}=-1,-2,\ldots,-k$, that is the conjugates of
the symmetric representations. This is precisely what we found in the previous section.

(b)  ${\mathfrak M}=SO(m)/SO(2)\times SO(m-2)$. The orbit takes the form
\EQ{
U\vec\gamma\cdot HU^{-1}=4\left(\BOmega\BOmega^\dagger-\BOmega^*\BOmega^t\right)\ ,
}
which means that
 \EQ{
\vec\gamma=8\vec\omega_1\ ,
}
where $\vec\omega_1$ is the highest weight of the defining $m$-dimensional vector representation of $SO(m)$. 
 In this case, $\ell=\frac18$ and using $\kappa=k/2$ we recover the quantization condition  \eqref{iuu2} and the Hilbert space consists of the symmetric representations of the $SO(m)$.
 
(c)  ${\mathfrak M}=Sp(m)/Sp(m-1)\times U(1)$. The orbit takes the form
\EQ{
U\vec\gamma\cdot HU^{-1}=4\left(\BOmega\BOmega^\dagger-J_m\BOmega^*\BOmega^tJ_m^{-1}\right)\ ,
}
which means that
 \EQ{
\vec\gamma=8\vec\omega_1\ ,
}
where $\vec\omega_1$ is the highest weight of the defining $2m$-dimensional vector representation of $Sp(m)$. 
 In this case, $\ell=\frac18$ and using $\kappa=k/2$ we recover the quantization condition  \eqref{iuu3} and the Hilbert space consists of the symmetric representations of the $Sp(m)$.

\subsection{The semi-classical spectrum}

Now we put together the results that we have established in the previous sections to consider the full semi-classical spectrum of elementary solitons.

For all the Grassmannians, AIII, BDI and CII, there is a spectrum solitons which have a quantized mass
\EQ{
M=\frac{4k}\pi\Big|m_1\sin\left(\frac{\pi{\EuScript N}}{2k}\right)\Big|\ ,\qquad|{\EuScript N}|=1,2,\ldots,k\ .
\label{cms}
}
with ${\EuScript N}>0$ only for BDI and CII. The states transform in rank-${\EuScript N}$ symmetric representations of $SU(n-p)$, $SO(n-p)$ and $Sp(n-p)$, respectively (or their conjugates for $\EuScript N<0$). For the case CII, the states also transform with respect to an $Sp(1)$ factor of $H$, in fact they transform in the product representation of $Sp(n-p)\times Sp(1)$ of the rank-$\EuScript N$ representations.
The states with lowest mass $|{\EuScript N}|=1$ have precisely the mass and quantum numbers of elementary perturbative excitations with mass $2|m_1|$ as appear in Table \ref{SpecTable}. 

The remaining set of excitations in Table \ref{SpecTable} can also be obtained from the solitons, if we suppose that the quantization condition on $\bar q$ in \eqref{iuu} extends to all the solitons and not just the subset with a non-trivial internal moduli space. At the moment our quantization technique does not apply to these solitons since they only have abelian charges and a trivial internal moduli space.  

For the general non-regular model described in Section 5.3, the solitons
transform in products of the rank-${\EuScript N}$ symmetric representation of $SU(d_a)$ and $SU(d_{a+1})$ (or their conjugates for ${\EuScript N}<0$) with masses
 \EQ{
M=\frac{4k}\pi\Big|(m_{a+1}-m_a)\sin\left(\frac{\pi{\EuScript N}}{2k}\right)\Big|\ ,\qquad{\EuScript N}=\pm1,\pm2,\ldots,\pm k\ .
}
Once again, the states at the bottom of the towers match the masses and quantum numbers of the perturbative states.
 
For AII, the soliton transforms under a pair $SU(2)\times SU(2)\subset H$ as an orbit of the form \eqref{efa}, {\it i.e.\/}
\EQ{
\frac{SU(2)}{U(1)}\times\frac{SU(2)}{U(1)}\ .
}
This identifies $\vec\gamma^{(a)}=\vec\gamma^{(a+1)}=4\vec\omega_1$, where $\vec\omega_1=1$ is the weight of the fundamental (spin $\frac12$) representation of $SU(2)$.\footnote{Note, that since $SU(2)$ has rank 1, the vector notation is actually redundant.}  Therefore, $l=\frac{1}{4}$, $\kappa=k$, and the quantization condition is as in \eqref{iuu}, which leads to the mass spectrum
\EQ{
M=\frac{4k}\pi\Big|(m_{a+1}-m_a)\sin\left(\frac{\pi{\EuScript N}}{2k}\right)\Big|\ ,\qquad{\EuScript N}=1,2,\ldots,k\ .
}
In this case, the states transform in the product of 
spin $\frac{\EuScript N}2$ representations of $SU(2)\times SU(2)$ and we only take ${\EuScript N}>0$ since the representations of $SU(2)$ are pseudo-real.

Finally, for the case of complex Grassmannians AIII with a non-regular $\Lambda$ the solitons have a mass
\EQ{
M=\frac{4k}\pi\Big|(m_{a+1}-m_a)\sin\left(\frac{\pi{\EuScript N}}{2k}\right)\Big|\ ,\qquad{\EuScript N}=\pm1,\pm2,\ldots,\pm k\ ,
} 
and transform in the product of the rank-$\EuScript N$ symmetric representations of $SU(d_a)\times SU(d_{a+1})$ (and conjugate representations for ${\EuScript N}<0$).

It is clear from all these examples that the soliton states at the base of the tower $|{\EuScript N}|=1$ have the same spectrum as the elementary perturbative excitations listed in Table~\ref{SpecTable}. This provides a strong check on our identification of solitons with the perturbative quanta.

For the case $F/G=SU(n+2)/U(n+1)\simeq\CP^{n+1}$, with $H=U(n)$, we have conjectured a form for the soliton S-matrix in \cite{Hollowood:2010rv}. In particular, the full quantum formula for the masses of the solitons, which transform in symmetric representations of $SU(n)$, generalizing \eqref{cms}, is
\EQ{
M=M_0\left|\sin\Big(\frac{\pi{\EuScript N}}{2k+n}\Big)\right|\ ,
}
where $M_0$ is an overall renormalized mass scale. It is tempting to conjecture that this mass formula applies to all the cases where $H$ has a single non-abelian factor with $n$ replaced by $h$, the dual Coxeter number of $H$. It is also interesting that the tower of solitons terminates at $|\EuScript N|=k$, in other words the spectrum only includes the symmetric representations which correspond to the symmetric representations of $\tilde H$ which are deformable to the quantum group $U_q(\tilde H)$ with deformation parameter
\EQ{
q=-\exp\Big(\frac{\pi i}{h+k}\Big)\ .
}
In the $F/G=SU(n+2)/U(n+1)$ case,  the conjectured S-matrix involves the trigonometric solution of the Yang-Baxter equation associated to this quantum group and it is natural to suppose that the quantum group structure generalizes to the other cases.

\section{Discussion}
\label{Discussion}

In this work, we have succeeded in constructing the soliton solutions for the class of SSSG theories associated to the compact type~I symmetric spaces. We have then proceeded to a semi-classical quantization of the solitons and found that the soliton spectrum includes the perturbative states of the theory. The fact that perturbative excitations and solitons can be identified follows from the fact that all states carrying charge have a kink-like behaviour, since in a gauge theory the Noether charge for global gauge transformations is a kink charge.

Part of the motivation for studying the SSSG theories is that they are related to the sigma model that describes string motion on the symmetric space $F/G$ by a process known as Pohlmeyer reduction.
In~\cite{Miramontes:2008wt}, the $F/G$ symmetric space sigma model is
formulated in terms of a group-valued field $\oldf  \in F$ and a gauge field $B_\mu\in{\mathfrak g}$ with a gauge symmetry 
\EQ{
\oldf \rightarrow \oldf  U^{-1}\>, \qquad
B_\mu \rightarrow U(B_\mu +\partial_\mu)U^{-1}\>,\qquad U\in G\>.
\label{GaugeTrans}
}
If the Lie group $F$ is simple, the nonlinear sigma model is defined by the Lagrangian
\begin{equation}
\LAG= -\frac{1}{2\kappa} \mathop{\rm Tr} \bigl(J_\mu J^\mu\bigr)\>,
\label{LagSM}
\end{equation}
where the current $J_\mu = \oldf ^{-1}\partial_\mu \oldf -B_\mu
\rightarrow U J_\mu U^{-1}$ is covariant under gauge transformations. 
The sigma model is then subjected to the Virasoro constraints which, up to conjugation, take the form
\EQ{
J_+=-\Lambda_+\ ,\qquad J_-=-\gamma^{-1}\Lambda_-\gamma\ ,
}
for constant elements $\Lambda_\pm\in{\mathfrak p}$ and $\gamma\in G$. The equations-of-motion of the sigma model can be written as the compatibility condition of the linear system \cite{Miramontes:2008wt}
\EQ{
\Big(\partial_++B_+-\Lambda_+\Big)f^{-1}=\Big(\partial_-+B_--\gamma^{-1}\Lambda_-\gamma\Big)f^{-1}=0\ .
}

Comparing with the SSSG equations-of-motion~\eqref{zcc} and~\eqref{LinProb}, we see that $\Lambda_\pm$ and $\gamma$ are identified with the same quantities in the SSSG equations, and the gauge fields become identified via
\EQ{
B_+=\gamma^{-1}\partial_+\gamma+\gamma^{-1}A_+\gamma\ ,\qquad B_-=A_-\ .
}
In addition, the group field of the sigma model is simply, along with \eqref{NormalDF} for comparison,
\EQ{
f=\Upsilon(z=1)^{-1}=\Upsilon_0(z=1)^{-1}\chi(z=1)^{-1}\ ,\qquad
\gamma=\chi(z=0)^{-1}\ .
\label{xll}
}
Consequently, it is a simple matter to construct the giant magnon solutions of the sigma model that correspond to the solitons in the SSSG equations. It is well known that these giant magnon solutions describe open strings, and this translates as the fact that the solitons of the SSSG system are kinks. The giant magnon system is not relativistically invariant, and this is manifested in \eqref{xll} because the spectral parameter $z$ transforms as $z\to e^\vartheta z$ under a Lorentz transformation, and so setting it to 1 breaks Lorentz symmetry explicitly.

In the sequel to this paper \cite{us2}, we extend our analysis to the SSSG theories related to Pohlmeyer reduction of the superstring on $AdS_5\times S^5$. The new element here is that the solitons have Grassmann and well as c-number collective coordinates.

\acknowledgments

TJH would like to acknowledge the support of STFC grant
ST/G000506/1.

\noindent 
JLM acknowledges the support of MICINN 
and FEDER (FPA2008-01838 and\break 
FPA2008-01177), Xunta de Galicia (Consejer\'\i a de Educaci\'on and INCITE09.296.035PR), and the 
Spanish Consolider-Ingenio 2010
Programme CPAN (CSD2007-00042).

\noindent
We would both like to
thank Arkady Tseytlin and Ben Hoare for discussions and comments on an earlier draft.
We also thank the Centro de F\'\i sica de Benasque for partial support while this work was in progress.

\startappendix

\Appendix{The Noether Current}
\label{App}

We will 
consider the expansion of the SSSG action in powers of $\phi$, where
$\gamma=e^\phi$, deduced by Hoare and Tseytlin in~\cite{Hoare:2009fs}, eq.~(2.14):\footnote{In the following expressions, we have already taken into account the relationship between our notation and the one used in~\cite{Hoare:2009fs}, summarized in footnote~1.}
\SP{
S[\gamma,A_\mu] = \frac{\kappa}{\pi}\int d^2 x\> \sum_{n=1}^{n} {\cal L}^{(n)}.
\label{HTAction}
}
Each term ${\cal L}^{(n)}$, of order $\phi^n$, is explicitly invariant under the gauge transformations
\EQ{
\phi \longrightarrow\phi^U= U\phi U^{-1},\qquad
A_\mu\longrightarrow A_\mu^U= U(A_\mu +\partial_\mu)U^{-1},\qquad U\in H.
\label{Gauge}
}
Moreover, all the terms ${\cal L}^{(n)}$ with $n\geq2$ are independent of the derivatives of $A_\mu$, like the original  action~\eqref{gWZW}. In contrast,
\SP{
{\cal L}^{(1)} &= -\Tr\Big(\left[D_+,D_-\right] \phi\Big) = -\Tr\Big(\left[\partial_+A_- -\partial_-A_+ +[A_+,A_-]\right] \phi\Big)\\[5pt]
&=-\Tr\Big(A_+\partial_-\phi -A_-\partial_+\phi +[A_+,A_-]\phi\Big)
-\partial_+\Tr\big(A_-\phi\big) +\partial_-\Tr\big(A_+\phi\big).
}
Therefore, the action~\eqref{HTAction} can be written as
\SP{
S[\phi,A_\mu]&= \frac\kappa{\pi}\int d^2 x\>\left( {\cal L}\big(\phi, \partial_\mu\phi, A_\mu\big) -\frac{1}{2}\epsilon^{\mu\nu}\partial_\mu \Tr\big(A_\nu \phi\big)\right),
\label{TopTerm}
}
such that
\EQ{
S[\phi^U,A_\mu^U]=S[\phi,A_\mu].
}

In particular, \eqref{TopTerm} is invariant under the global gauge transformations
\EQ{
\phi \longrightarrow U\phi U^{-1},\qquad
A_\mu\longrightarrow  UA_\mu U^{-1},\qquad U\in H
\label{Global}
}
and the corresponding Noether current can be found following standard means by considering a local version of the transformations with $U=U(t,x)$. Gauge invariance is the statement that 
\SP{
S[U\phi U^{-1},UA_\mu U^{-1}] =S[\phi,A_\mu + U^{-1}\partial_\mu U].
\label{NotGlobal}
}
Consider now an infinitesimal transformation $U= 1+u+\cdots$, and assume that $\phi$ and $A_\mu$ satisfy the equations of motion with respect to $A_\mu$; namely $\partial{\cal L}/\partial A_\mu =0$ (the equations-of-motion of $\phi$ are not required). Then, to linear order
\SP{
\delta_uS=S[\phi,A_\mu + \partial_\mu u] -S[\phi,A_\mu] =
-\frac{\kappa}{2\pi}\int d^2 x\> \epsilon^{\mu\nu}\partial_\mu \Tr\big(\phi\partial_\nu u\big),
\label{InfGauge}
}
which leads to the on-shell expression for the Noether current
\EQ{
{\cal J}_\mu \approx  \epsilon_{\mu\nu} \partial^\nu \phi^\perp\ .
}
This current is topological, in agreement with the general arguments of  \cite{Julia:1998ys,Silva:1998ii,Julia:2000er} (and references therein). Moreover, since
\EQ{
\gamma(\pm\infty)= e^{\phi(\pm\infty)},
}
and $\phi(\pm\infty)\in{\mathfrak h}$ commute, the Noether charge is precisely the kink charge
\EQ{
\int_{-\infty}^{+\infty} dx\> {\cal J}^0 \approx -\phi(\infty)+\phi(-\infty)= q_0\ .
}

In order to describe the semi-classical dynamics of the solitons we will need the action for a soliton conjugated by a time-dependent transformation in $H$. It is important to understand that this is not a gauge transformation since the gauge field remains at $A_\mu=0$. However, we can use the gauge invariance condition \eqref{NotGlobal} to write, 
with $U=U(t)$,
\SP{
S[U\phi U^{-1},0] =S[\phi,U^{-1}\partial_\mu U]\ .
}
From \eqref{TopTerm}, we find, therefore,
\SP{
S[U\phi U^{-1},0] -S[\phi,0]= -\frac{\kappa}{2\pi}\int d^2x\> \epsilon^{\mu\nu} \partial_\mu\Tr\big(U^{-1}\partial_\nu U\> \phi\big) +\cdots
}
where the dots represent terms of higher order in $U^{-1}\partial_\mu U$ which come from the first term in \eqref{TopTerm}:
\EQ{
{\cal L}(\phi,\partial_\mu\phi,A_\mu=U^{-1}\partial_\mu U)\ .
} 
Concentrating on the 
linear term, which is the dominant one in the semi-classical limit,
since $U=U(t)$ we can perform the integral over $x$ and write the result as an effective quantum mechanical action: 
\SP{
S_\text{eff}[U]=S[U\phi U^{-1},0] -S[\phi,0]= \frac{\kappa}{2\pi}\int dt\> \Tr\Big(U^{-1}\frac{dU}{dt}\> q_0\Big) +\cdots\ ,
\label{TheMissingLink}
}
with the corrections being quadratic in $U^{-1}\dot U$.

In the last part of this appendix, we develop a more general understanding of the Noether current and derive an off-shell expression for it. One of the subtleties in the SSSG
theories is that the fields are non-trivial at $x=\pm\infty$ and so the WZ term and its variation requires careful treatment. 
One way to unambiguously define the action is, as in~\cite{Hoare:2009fs}, to use the condition of gauge invariance to pin down the expansion of the WZ term in terms of $\phi$. The prescription amounts to taking an action
\SP{
S=&S_\text{WZW}[\gamma] -\frac\kappa{\pi} \int d^2x\> \Tr\Big[A_+ \partial_-\gamma\gamma^{-1} -A_-\gamma^{-1}\partial_+\gamma\\[5pt]
&
-A_+\gamma A_-\gamma^{-1} + A_+A_-- \Lambda\gamma^{-1}\Lambda\gamma+\frac{1}{2} \epsilon^{\mu\nu}\partial_\mu\big(A_\nu \phi\big)\Big]
}
with a suitable prescription for the WZ term. Then, the off-shell version of~\eqref{InfGauge} is
\SP{
\delta_u S=& \frac\kappa{\pi}\int d^2 x\> \Tr\Big[\partial_+u\> (\gamma\partial_-\gamma^{-1}+\gamma A_-\gamma^{-1} -A_-)
\\[5pt]
&+\partial_-u\> (\gamma^{-1}\partial_+\gamma+\gamma^{-1}A_+\gamma -A_+)-\frac12\epsilon^{\mu\nu}\partial_\mu\big(\phi\partial_\nu u\big)\Big],
}
and the off-shell Noether current follows as
\SP{
{\cal J}_\pm=(\gamma^{\mp1}\partial_\pm\gamma^{\pm1}+\gamma^{\mp1}A_\pm\gamma^{\pm1} -A_\pm)^\perp\mp\partial_\pm\phi^\perp\ .
\label{ncor}
}
An important aspect of this equations is that it is valid for any choice of the gauge fixing prescription.

\end{document}